\documentclass[fleqn,usenatbib]{mnras}

\usepackage{newtxtext,newtxmath}

\usepackage[T1]{fontenc}
\usepackage{ae,aecompl}

\usepackage{graphicx}	
\usepackage{amsmath}	
\usepackage{amssymb}	
\usepackage{wasysym}
\usepackage{tabularx}

\title[ETG density profile evolution]{Early-type galaxy density profiles from IllustrisTNG: \\ II. Evolutionary trend of the total density profile}

\author[Y. Wang et al.]{Yunchong Wang$^{1,2,3}$\thanks{E-mail: \url{ycwang15@mit.edu}},
Mark Vogelsberger$^{1}$,
Dandan Xu$^{4}$,
Xuejian Shen$^{5}$,
\newauthor
Shude Mao$^{2, 6}$,
David Barnes$^{1}$,
Hui Li$^{1}$,
Federico Marinacci$^{7}$,
Paul Torrey$^{8}$,
\newauthor
Volker Springel$^{9}$
and Lars Hernquist$^{10}$
\\
$^{1}$Kavli Institute for Astrophysics and Space Research, Department of Physics, MIT,  Cambridge, MA 02139, USA\\
$^{2}$Department of Astronomy and Tsinghua Center for Astrophysics, Tsinghua University, Beijing 100084, China\\
$^{3}$Department of Physics, Tsinghua University, Beijing, 100084, China\\
$^{4}$Institute for Advanced Studies and Tsinghua Center for Astrophysics, Tsinghua University, Beijing, 100084, China\\
$^{5}$The Division of Physics, Math and Astronomy, California Institute of Technology, Pasadena, CA 91125, USA\\
$^{6}$National Astronomical Observatories, Chinese Academy of Sciences, Beijing, 100012, China\\
$^{7}$Department of Physics $\&$ Astronomy, University of Bologna, via Gobetti 93/2, 40129 Bologna, Italy\\
$^{8}$Department of Astronomy, University of Florida, 211 Bryant Space Sciences Center, Gainesville, FL 32611, USA\\
$^{9}$Max-Planck-Institut f\"{u}r Astrophysik, Karl-Schwarzschild-Str. 1, D-85748, Garching, Germany\\
$^{10}$Harvard-Smithsonian Center for Astrophysics, 60 Garden Street, Cambridge, MA 02138
}

\date{Accepted 2019 October 11. Received 2019 October 9; in original form 2019 June 25}

\pubyear{2019}

\begin{document}
\label{firstpage}
\pagerange{\pageref{firstpage}--\pageref{lastpage}}
\maketitle

\begin{abstract}
We study the evolutionary trend of the total density profile of early-type galaxies (ETGs) in IllustrisTNG. To this end, we trace ETGs from $z=0$ to $z=4$ and measure the power-law slope $\gamma^{\prime}$ of the total density profile for their main progenitors. We find that their $\gamma^{\prime}$ steepen on average during $z\sim4-2$, then becoming shallower until $z=1$, after which they remain almost constant, aside from a residual trend of becoming shallower towards $z=0$. We also compare to a statistical sample of ETGs at different redshifts, selected based on their luminosity profiles and stellar masses. Due to different selection effects, the average slopes of the statistical samples follow a modified evolutionary trend. They monotonically decrease since $z=3$, and after $z\approx 1$, they remain nearly invariant with a mild increase towards $z=0$. These evolutionary trends are mass-dependent for both samples, with low-mass galaxies having in general steeper slopes than their more massive counterparts. Galaxies that transitioned to ETGs more recently have steeper mean slopes as they tend to be smaller and more compact at any given redshift. By analyzing the impact of mergers and AGN feedback on the progenitors' evolution, we conjecture a multi-phase path leading to isothermality in ETGs: dissipation associated with rapid wet mergers tends to steepen $\gamma^{\prime}$ from $z=4$ to $z=2$, whereas subsequent AGN feedback (especially in the kinetic mode) makes $\gamma^{\prime}$ shallower again from $z=2$ to $z=1$. Afterwards, passive evolution from $z=1$ to $z=0$, mainly through gas-poor mergers, mildly decreases $\gamma^{\prime}$ and maintains the overall mass distribution close to isothermal.
\end{abstract}




\section{Introduction}
\label{sec: 1}

Early-type galaxies (hereafter, ETGs) are the end products of galaxy formation and evolution. It is thought that they first formed by accreting cold gas from cosmic filaments above $z = 3$~\citep{1977MNRAS.179..541R,1991ApJ...379...52W,2005MNRAS.363....2K,2009Natur.457..451D,2011MNRAS.414.2458V,2015MNRAS.448...59N}, and subsequently quenched their star formation activities by Active Galactic Nuclei (hereafter, AGN) feedback from the central supermassive black hole~\citep{1998A&A...331L...1S,2003ApJ...595..614W,2003ApJ...596L..27K,2005MNRAS.361..776S,2005Natur.433..604D,2012ARA&A..50..455F,2013ARA&A..51..511K}, and evolving passively since $z \approx 2$. Their assembly history is crucial for constraining the structure formation theory under the most widely tested $\Lambda$-Cold Dark Matter (CDM) cosmology model~(e.g. \citealt{1978MNRAS.183..341W,1984Natur.311..517B,1994MNRAS.271..781C}). The typical morphology of these galaxies is elliptical or lenticular, and they have been observed and studied in detail through methods such as strong and weak gravitational lensing~\citep{2006MNRAS.368..715M,2007ApJ...667..176G,2012ApJ...744...41B,2015ApJ...803...71S,2017ApJ...851...48S}, stellar kinematics~\citep{2011MNRAS.414.2923K,2011MNRAS.414..888E,2012ApJ...748....2D,2014MNRAS.439..659N,2014MNRAS.444.3357N,2016ARA&A..54..597C,2017MNRAS.464..356V}, and stellar dynamic modelling ~\citep{2009MNRAS.393..491C,2011MNRAS.413..813C,2013MNRAS.432.1709C,2014MNRAS.445..162T,2017ApJ...838...77L,2017MNRAS.467.1397P,2018MNRAS.473.1489L,2018MNRAS.476.1765L,2018ApJ...863L..19L}. 

The total matter density profile described by a power-law model $\rho(r) \propto r^{-\gamma^{\prime}}$ is a first-order but important approximation of an ETG. Especially for the baryon dominated inner region, $\gamma^{\prime}$ serves as an indicator of the compactness of the matter distribution within a few effective radii. It is well known from observations of strong and weak lensing of high-redshift ETGs~\citep{2006ApJ...649..599K,2007ApJ...667..176G,2009ApJ...703L..51K,2009MNRAS.399...21B,2011MNRAS.415.2215B,2010ApJ...724..511A,2018MNRAS.480..431L,2018MNRAS.475.2403L}, X-ray studies~\citep{2006ApJ...646..899H,2010MNRAS.403.2143H} and local dynamically modeled ETGs~\citep{2014MNRAS.445..115T,2015ApJ...804L..21C,2016MNRAS.460.1382S,2017MNRAS.467.1397P,2018MNRAS.476.4543B} that the power-law slope $\gamma^{\prime}$ is close to 2, resembling a self-gravitating isothermal collisional ideal gas sphere. Since neither the stellar nor the dark matter component follow such a density profile, this combined effect is also known as the `bulge-halo conspiracy'. Current strong lensing observations out to redshift $z=1$, i.e. SLACS~\citep{2011ApJ...727...96R}, SL2S~\citep{2013ApJ...777...98S}; S4TM~\citep{2017ApJ...851...48S}; BELLS~\citep{2012ApJ...757...82B,2018MNRAS.480..431L}; Herschel-ATLAS~\citep{2014MNRAS.440.2013D}, consistently show a mild steepening trend of the power-law slope towards lower redshift.

Since it is infeasible to observe the evolution of individual galaxies over time, theoretical approaches focusing on understanding the formation of ETGs have made use of numerical simulations to trace the evolution of individual galaxies. Through zoom-in and cosmological simulations, a consensus has emerged between these simulations that the formation of ETGs proceeds through two phases, where galaxies first go through dissipative gas-rich wet mergers followed by in-situ star formation bursts at redshifts above $z\approx2$, and then evolve towards low redshift through non-dissipative gas-poor dry mergers ~\citep{2007ApJ...658..710N,2008MNRAS.384....2G,2009ApJ...703.1531N,2009ApJ...706L..86N,2009ApJ...691.1168H,2010ApJ...725.2312O,2012ApJ...754..115J,2013MNRAS.428.3121M,2013ApJ...766...71R,2015MNRAS.450.4486F,2015MNRAS.449..361W,2016MNRAS.456.1030W,2016MNRAS.458.2371R}. However regarding the redshift evolution of ETGs' total power-law density slopes, no consensus has been reached neither among different cosmological hydrodynamic simulations nor between the simulations and observations, despite the many advances in cosmological simulations~\citep{2019arXiv190907976V}. While the Magneticum pathfinder simulation~\citep{2017MNRAS.464.3742R} and the Illustris Simulations~\citep{2017MNRAS.469.1824X} produce shallower total density profile with time, the Horizon-AGN simulations~\citep{2019MNRAS.483.4615P} produce steeper total density profile with time, in better agreement with the redshift evolution trend found in observations. However, the latter simulation has smaller slope values compared to the former two, which are closer to the observed slope values due to different implementation of feedback models etc. Apart from cosmological simulations, dedicated zoom-in simulations~\citep{2009ApJ...690..802J,2012ApJ...754..115J,2013ApJ...766...71R} have revealed that dry mergers that dominate the passive evolution of ETGs below $z\approx2$ could make the total density profile shallower than isothermal \citep{2012MNRAS.425.3119H,2013MNRAS.429.2924H,2017MNRAS.464.3742R}. The inclusion of wet mergers is also crucial for reconciling the simulated redshift evolution trend of the slope with strong-lensing observations~\citep{2014ApJ...786...89S}. 

In \citet{2018arXiv181106545W} (Paper I hereafter) we studied a sample of ETGs from a state-of-the-art cosmological hydrodynamic simulations, the IllustrisTNG Simulations\footnote{\url{http://www.tng-project.org}}~\citep{2018MNRAS.480.5113M,2018MNRAS.477.1206N,2018MNRAS.475..624N,2018MNRAS.475..648P,2018MNRAS.475..676S}. These galaxies show correlations between the total power-law slope with galaxy size, stellar mass, surface density and central dark matter fraction that are broadly consistent with both local and distant observations. The total power-law slope also correlates with the in-situ formed stellar mass ratio and redshift, in line with previous simulation studies~\citep{2017MNRAS.469.1824X,2017MNRAS.464.3742R,2018MNRAS.476.4543B}. Although a slightly negative total density slope/central velocity dispersion correlation and a shallower slope with time are in tension with observational trends, the uncertainties and systematics in observations may have obscured the true trends~\citep{2016MNRAS.456..739X,2017MNRAS.469.1824X,2018MNRAS.474.3403T}. In this work, we will hence investigate the redshift evolution of the density profiles of $z = 0$ ETG progenitors, and statistical ETG samples at higher redshifts selected based on their morphology. We will analyze the dependence of the total power-law slope on stellar mass, quenching time, merger events, AGN feedback history etc.

This paper is organized as follows: in Section~\ref{sec: 2} we introduce our methodology to analyze the total density profile, merger history, AGN feedback, and other galaxy properties of the selected ETGs; in Section~\ref{sec: 3} we present the density profile evolution of both ETG progenitors and statistical ETG samples in different stellar mass bins, and the variance of the total power-law slope at $z = 0$ with respect to different quenching time; in Section~\ref{sec: 4}, we study the effects of merger events and AGN feedback history on the evolution of the total power-law slopes of ETG progenitors; in Section~\ref{sec: 5} we summarize the major aspects involved in the formation scenario of IllustrisTNG ETGs and their implications for galaxy formation processes. In this work, the Planck $\Lambda$CDM cosmology~\citep{2016A&A...594A..13P} has been used in all analyses, which has also been the cosmology model assumed in the IllustrisTNG Simulations; i.e., $h = 0.6774$, $\Omega_{\mathrm{m}} = 0.3089$, $\Omega_{\mathrm{\Lambda}} = 0.6911$, $\Omega_{\mathrm{b}} = 0.0486$, and $\sigma_{\mathrm{8}} = 0.8159$.

\section{Methodology}
\label{sec: 2}

\subsection{The simulation}
\label{sec: 2.1}

The IllustrisTNG Simulations (IllustrisTNG hereafter, see  \citealt{2018MNRAS.480.5113M,2018MNRAS.477.1206N,2018MNRAS.475..624N,2018MNRAS.475..648P,2018MNRAS.475..676S}) are a set of state-of-the-art magneto-hydrodynamic cosmological simulations consisting of 3 box sizes, i.e. TNG300, TNG100, and TNG50 (with $302.6\,\mathrm{Mpc}$, $110.7\,\mathrm{Mpc}$, and $51.7\,\mathrm{Mpc}$ box length respectively). We select galaxies from the highest resolution version of the TNG100 simulation, which has dark matter and baryon mass resolution of $m_{\mathrm{DM}} = 7.5\times 10^{6}\mathrm{M}_{\mathrm{\astrosun}}$ and $m_{\mathrm{baryon}} = 1.4\times10^{6} \mathrm{M}_{\mathrm{\astrosun}}$. TNG100 and TNG300 simulation data are available for public access~\citep{2019ComAC...6....2N}.

IllustrisTNG inherits many successful aspects of the Illustris Simulations~\citep{2014MNRAS.444.1518V,2014Natur.509..177V,2014MNRAS.445..175G,2015A&C....13...12N,2015MNRAS.452..575S}, which was evolved using the advanced moving-mesh hydrodynamics code \textsc{arepo}~\citep{2010MNRAS.401..791S}. The IllustrisTNG physical model has made major improvements in the AGN feedback and galactic wind models~\citep{2017MNRAS.465.3291W,2018MNRAS.473.4077P} over the original Illustris models~\citep{2013MNRAS.436.3031V,2014MNRAS.438.1985T}. The baryonic version of the IllustrisTNG suite reproduced many observational relations, including the evolution of the mass-metallicity relation~\citep{2017arXiv171105261T,2018MNRAS.477L..16T}, the galaxy-color bimodality in the Sloan Digital Sky Survey~\citep{2018MNRAS.475..624N}, the fraction of dark matter within galaxies at $z=0$~\citep{2018MNRAS.481.1950L}, the galaxy size-mass relation evolution~\citep{2018MNRAS.474.3976G}, the cool-core structure in galaxy clusters~\citep{2018MNRAS.481.1809B}, and the intra-cluster metal distribution in galaxy clusters~\citep{2018MNRAS.474.2073V}. The many successes in producing galaxy relations consistent with observations lends credibility to utilizing the simulated IllustrisTNG galaxy and galaxy cluster populations for theoretical purposes. 

\subsection{The SUBLINK merger tree}
\label{sec: 2.2}

A central aspect of this work is that we trace the individual evolutionary tracks of an ETG sample selected at $z=0$ up to $z=4$, which is complementary to previous observational and theoretical studies. We climb up the `main progenitor branch' (MPB) of selected IllustrisTNG ETGs in the galaxy (baryonic) version of the \textsc{sublink} merger tree~\citep{2015MNRAS.449...49R}. The `main progenitor branch' of the merger tree for a galaxy is determined by the \textsc{sublink} algorithm which follows the subhalo branch with the `most massive history'~\citep{2007MNRAS.375....2D}. We will also identify mergers through the merger tree, since the assembly of the ex-situ formed stellar population is dominated by major and minor mergers~\citep{2016MNRAS.458.2371R}, and correlates with the total power-law slope~\citep{2017MNRAS.464.3742R,2018MNRAS.476.4543B}. Following the approach of \citet{2015MNRAS.449...49R}, all galaxy properties of the in-falling galaxy that merged into the main progenitors will be calculated at the snapshot in which it had the largest value of stellar mass enclosed within twice its stellar half mass radius. We define the merger stellar mass ratio $\mu_{\ast}$ as the ratio of the stellar mass between the less massive progenitor over the more massive progenitor, and we only consider merger events with $\mu_{\ast}\geqslant 0.01$. We further define mergers with $\mu_{\ast}\geqslant 0.25$ as major mergers, and mergers with $\mu_{\ast} < 0.25$ as minor mergers.

Besides tracing the evolution of the total density profile and merger events along the MPB, we also record the stellar mass (enclosed within central $30\,\mathrm{kpc}$), AGN feedback (kinetic and thermal mode), total cold gas fraction, and in-situ formed stellar mass ratio. For the statistically selected sample of ETGs at different redshifts (namely $z = 0$, 0.1, 0.2, 0.3, 0.4, 0.5, 0.7, 1.0, 1.5, 2.0, and 3.0), we define their stellar masses as the total stellar mass enclosed within the central $30\,\mathrm{kpc}$ of these galaxies. The stellar mass for the traced progenitor sample is approximated by the stellar mass enclosed within twice the 3D stellar half mass radius which is sufficiently large to cover most of the stellar mass in progenitor galaxies while reducing the pollution from satellites.

We point out that the \textsc{sublink} algorithm may occasionally misidentify the MPB during a galaxy merger causing a `halo-switch' issue that could temporarily decrease the progenitor stellar mass by orders of magnitude. To overcome this issue, we identify the `dips' in the stellar mass evolution tracks of MPB galaxies and neglect the corresponding snapshots. The criterion we use to identify these `dips' is:
\begin{equation}
M_{\ast,\,\mathrm{i}}^{2} < 0.25 \times M_{\ast,\,\mathrm{i-1}} \times M_{\ast,\,\mathrm{i+1}}\,,
\end{equation}
where $M_{\ast,\,\mathrm{i-1}}$, $M_{\ast,\,\mathrm{i}}$, and $M_{\ast,\,\mathrm{i+1}}$ are the stellar masses of the main progenitors in the $i-1\,$th, $i\,$th and $i+1\,$th snapshot, and $i$ ranges from 22 to 98 corresponding to the redshift range $0$ to $4$. After neglecting these snapshots, we interpolate all physical quantities that we trace as the mean of those quantities in the snapshots just before and after the snapshot we neglected. We note that this smoothing criterion is sufficient to remove most `halo-switch' events, but it occasionally ($\lesssim 10\%$ of all events) leaves out such events if the masses of the three consecutive main progenitors in a `halo-switch' event are just below the criterion margin.

\subsection{Galaxy selection and analysis}
\label{sec: 2.3}

Galaxies are identified as gravitationally-bound systems by \textsc{subfind}~\citep{2001MNRAS.328..726S,2009MNRAS.399..497D} in IllustrisTNG. In \citet{2017MNRAS.469.1824X} and \citet{2018arXiv181106545W}, the selection and classification of the simulated galaxies have been described in detail, and we apply the same methods in this work. We derive the optical luminosity of the galaxy with the SPS model \textsc{galaxev}~\citep{2003MNRAS.344.1000B} based on the age and metallicity of every stellar particle in the galaxy, and apply a basic dust attenuation treatment that varies with the viewing angle. Projected along the three principal axes (X, Y, and Z directions) of the simulation box, every galaxy's radial luminosity profile is calculated and fitted with a de Vaucouleurs, an exponential, and a bulge-disk 2-component radial luminosity model. The fitting procedure is conducted in a minimum-$\chi^{2}$ fashion, and we consider the model which gives a lower minimum-$\chi^{2}$ a better model. Galaxies better fitted by the de Vaucouleurs model than the exponential model, and which have a large bulge-to-total ratio ($>50\%$) in the best-fit two-component model in all three independent projections are classified as early-type. 

We select statistical samples of ETGs that have stellar mass $10^{10.7}\mathrm{M_{\astrosun}}\leqslant M_{\ast}\leqslant 10^{11.9}\mathrm{M_{\astrosun}}$ enclosed within the central $\mathrm{30\,kpc}$ region. They are well-resolved and resemble the observed ETGs~\citep{2010ApJ...724..511A,2011ApJ...727...96R,2013ApJ...777...98S}. In addition to the above criteria, we add a constraint on the S$\mathrm{\acute{e}}$rsic index $n$ of selected galaxies, $n\geqslant2$, to ensure a more robust ETG classification. This results in 491 IllustrisTNG ETGs at $z=0$, and different numbers for different redshifts (see \ref{sec: 3.1} for details) that make up our statistical sample of ETGs ranging from $z = 0$ to $z=3$.

For the progenitor tracing sample, we choose galaxies from the 491 $z=0$ IllustrisTNG ETGs in the statistical sample that were also classified as ETGs in the $z=1$ snapshot. This consideration is meant to mimic the passively evolving ETG population already quenched since $z=1$ as predicted by the two-phase formation scenario, and results in 165 ETGs for which we trace back various galaxy properties along their merger tree of the main progenitor branch from $z=0$ to $z=4$. We will also explore the selection effects induced by artificially choosing different quenching time for the $z=0$ sample total power-law density slope in Section \ref{sec: 3.2}.

The total power-law density slope, $\gamma^{\prime}$ of the total density profile in the form of $\rho(r)\propto r^{-\gamma^{\prime}}$, is calculated over the range $0.4\,R_{1/2}$ to $4\,R_{1/2}$, where $R_{1/2}$ is the 3D stellar half mass radius of IllustrisTNG ETGs. This was also the radial range for the total power-law slope we adopted to investigate various galaxy correlations in Paper I. Assuming spherical symmetry, we perform a linear fit (with equal radial bin weighting) to $\log\rho(r)-\log r$ within this radial range and define the best linear fit slope (minimum $\chi^{2}$) as the total density power-law slope $\gamma^{\prime}$ for each of our galaxies. 

Since the main progenitors of $z=0$ ETGs traced to high redshifts are not guaranteed to be well-resolved (having $\approx 10^{4}$ particles), we adopt fewer radial bins to suppress the Poisson noise in those low-resolution galaxies, i.e. 100 logarithmic radial bins for ETGs with total particle number $N \geqslant 5000$, 30 logarithmic radial bins for ETGs with particle number $1000 \leqslant N < 5000$, and we ignore ETG progenitors with particle number $N < 1000$. This ensures reasonably resolved total density profile power-law slope fits out to $z=4$ for the MPB sample, and the statistical ETG sample stellar mass lower bound of $\mathrm{5\times 10^{10}\, M_{\astrosun}}$ guarantees the 100 logarithmic radial bins criterion automatically.

\section{The evolution history of the total density profile}
\label{sec: 3}

In this section, we present the mean evolution path of the IllustrisTNG ETG total power-law density slope traced along the main progenitor branch of the \textsc{sublink} merger tree and in statistical samples selected at different redshifts. We will also show the dependence of the slope evolution on stellar mass and quenching redshift.

\subsection{Sample selection and stellar mass evolution}
\label{sec: 3.1}

\begin{figure}
\includegraphics[width=\columnwidth]{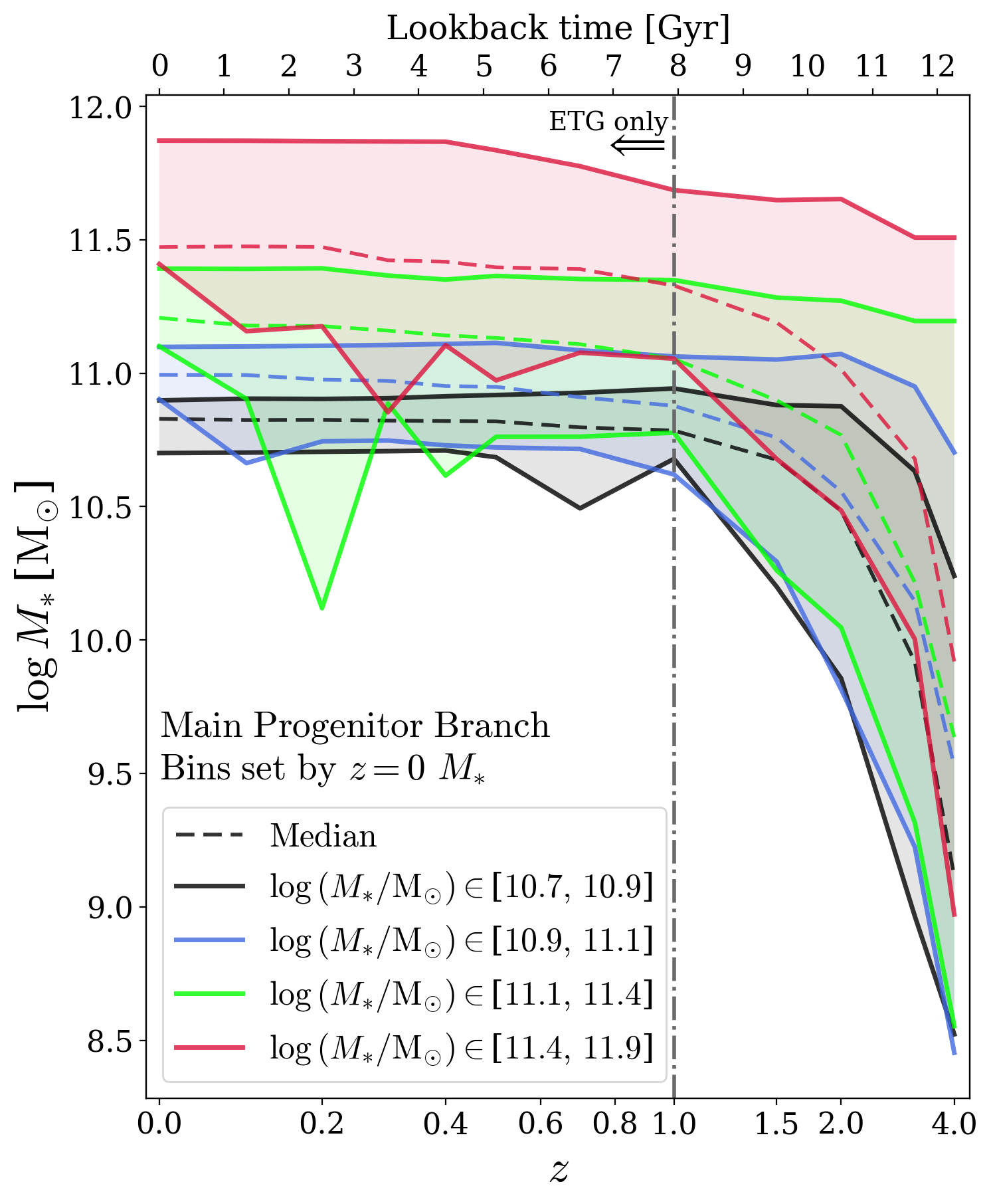}
\caption{The evolution of the total stellar mass within the central $30\,\mathrm{kpc}$ of the main progenitors traced by the merger tree from $z=0$ to $z=4$ (same redshift range as the statistical ETG sample). The four stellar mass bins of sources are divided by their $z = 0$ stellar masses. The median of the stellar mass distribution in each bin is shown by the dashed curves, and the maximum/minimum values are marked by the solid curves of the same color. The median value does not increase much in the four mass bins from $z=1$ to $z=0$, but increased rapidly before $z=1$. This reflects our selection criteria which require the progenitors to quench and become ETGs by $z=1$, and evolve passively afterwards.}
\label{fig:m_ev}
\end{figure}

\begin{figure}
\includegraphics[width=\columnwidth]{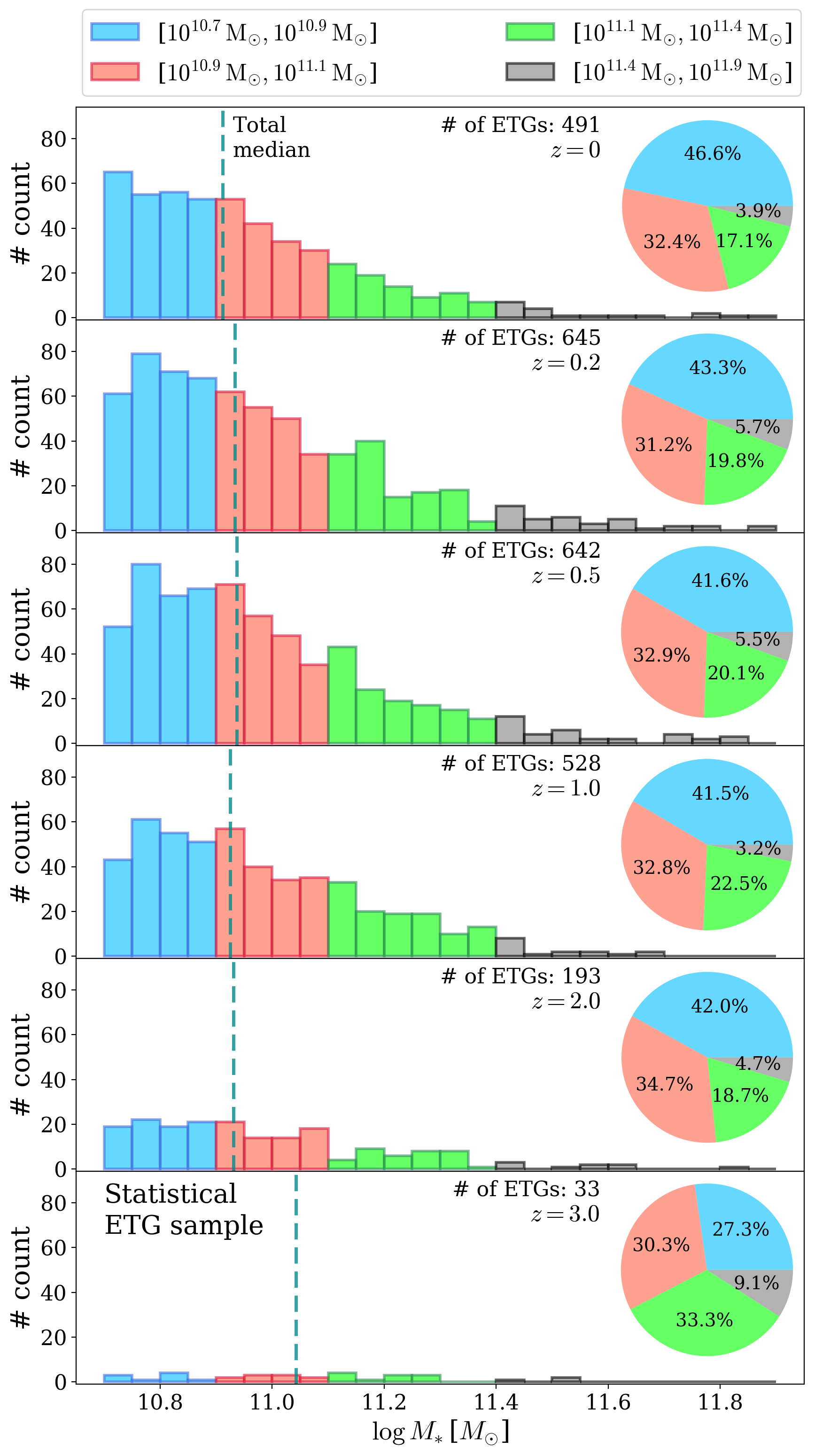}
\caption{The evolution of the total stellar mass distribution for the statistical ETG samples selected at redshifts 0, 0.2, 0.5, 1, 2, and 3. The distribution of the four stellar mass bins $\mathrm{log}\,(M_{\ast}/\mathrm{M_{\astrosun}})\in\,$[$10.7$, $10.9$], [$10.9$, $11.1$], [$11.1$, $11.4$], and [$11.4$, $11.9$] are shown by the blue, red, green and black histograms, respectively. The median stellar mass of each distribution is given by the dashed line in each subplot, and the total number of ETGs in each snapshot is also labeled. The pie chart to the right of each subplot shows the proportion of ETGs in each of the four stellar mass bins with the same color reference as the histogram. }
\label{fig:stats_mev}
\end{figure}

The \textbf{MPB tracing sample}, is composed of a selected set of 165 ETGs at $z=0$ together with their MPBs all the way to $z=4$. We require that the MPBs of the selected present-day ETGs must also be classified as ETGs at $z=1$. The progenitors of this sample above $z=1$ are not necessarily ETGs. Note that the reason that we require such a specific redshift upper limit on type transition is two-fold: (1) the majority of the most massive ETGs within the given mass range is quenched around $z=1$ (see Figure 7 of \citealt{2018MNRAS.474.3976G}); (2) massive ETGs that are observed to high redshifts seem to have evolved little since $z=1$~(e.g. \citealt{ 2006ApJ...649..599K,2009ApJ...703L..51K,2010ApJ...724..511A,2011ApJ...727...96R,2012ApJ...757...82B,2013ApJ...777...98S,2014MNRAS.440.2013D,2017ApJ...840...34S,2018MNRAS.480..431L}). In Section~\ref{sec: 3.4}, we vary this redshift limit to see how this would affect the evolutionary trend of a given selected sample, especially for the effects of lower mass galaxies quenching and joining in the ETG population at $z=0$ later than their higher mass counterparts. 

In order to study the mass dependence of the density slope evolution, we divide galaxies in all MPBs into 4 mass bins based on their $z=0$ stellar mass enclosed within the central $30\,\mathrm{kpc}$ region, i.e. $\mathrm{log}\,(M_{\ast}/\mathrm{M_{\astrosun}})\in\,$[$10.7$, $10.9$], [$10.9$, $11.1$], [$11.1$, $11.4$], and [$11.4$, $11.9$]. Note that their progenitor stellar masses will in general not remain in these 4 stellar mass bins as we trace them along the MPB to higher redshifts. 

We present the stellar mass evolution in the 4 stellar mass bins for the main progenitor tracing from $z=0$ to $z=4$ in Fig.~\ref{fig:m_ev}. For the main progenitors, the medians of the stellar mass ranges in the 4 mass bins are almost constant below $z=1$, but drop rapidly with time above $z=1$. This is typical for the two-phase formation scenario of massive galaxies, where active in-situ star formation above $z\approx 2$ that rapidly increases the stellar mass is followed by accretion of ex-situ stellar populations below $z\approx2$~~\citep{2007ApJ...658..710N,2008MNRAS.384....2G,2009ApJ...703.1531N,2009ApJ...706L..86N,2009ApJ...691.1168H,2010ApJ...725.2312O,2012ApJ...754..115J,2013MNRAS.428.3121M,2013ApJ...766...71R,2015MNRAS.450.4486F,2015MNRAS.449..361W,2016MNRAS.456.1030W,2016MNRAS.458.2371R}. Note that at $z=3$ and $z=4$, the least massive galaxies are resolved by $\gtrsim 10000$ and $\gtrsim 4000$ particles in total.

The \textbf{statistical ETG sample} is a set of ETGs selected at $z=0,\ 0.1,\ 0.2,\ 0.3,\ 0.4,\ 0.5,\ 0.7,\ 1.0,\ 1.5,\ 2,$ and $3$ purely based on their luminosity profiles and stellar masses at each of these redshifts as described in Section~\ref{sec: 2.3} (galaxies at higher redshifts are not necessarily progenitors of those at lower redshifts). To study the mass dependence, we also divide the statistical ETG sample into 4 mass bins based on their stellar masses at all selected redshifts. We show the stellar mass and number evolution in the 4 stellar mass bins for the statistically selected ETG sample from $z=0$ to $z=3$ in Fig.~\ref{fig:stats_mev}. For succinctness we only show redshifts 0, 0.2, 0.5, 1, 2, 3 for the mass and ETG number evolution. As shown in the figure, the number of galaxies in the lowest mass bin decreases with increasing redshift, so does the median mass at different redshifts. The steep stellar mass function towards lower redshift reflects the increasing dominance of low-mass ETGs with time. Above $z=2$, the total number of ETGs within the required stellar mass range drops significantly. We therefore do not consider any ETG sample beyond $z>3$. 

\subsection{The total power-law density slope evolution}
\label{sec: 3.2}

\begin{figure}
\includegraphics[width=\columnwidth]{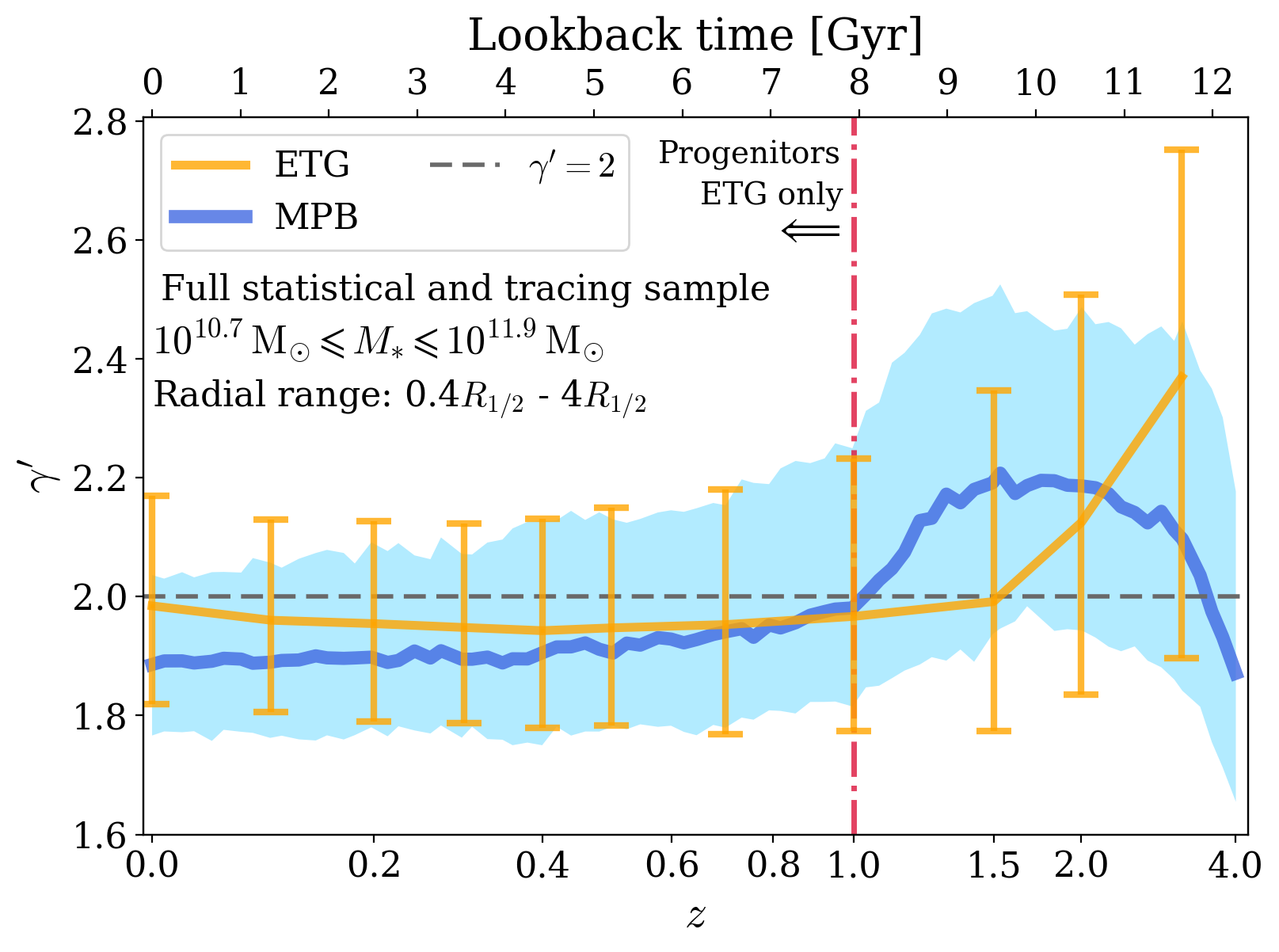}
\caption{The evolutionary trend of the total power-law density slope $\gamma^{\prime}$ with redshift. The MPB tracing sample is shown in blue, with the solid curve denoting the median and the shaded region denoting the [$16\%$, $84\%$] distribution of $\gamma^{\prime}$, traced from $z=0$ to $z=4$. The stellar masses within the central $30\,\mathrm{kpc}$ of the sample selected at $z=0$ are in the range of $\mathrm{log}\,(M_{\ast}/\mathrm{M_{\astrosun}})\in\,$[$10.7$, $11.9$]. Their selection enforced them to have already become ETGs morphologically by $z=1$, shown by the red dashed dotted line. The distribution of the statistical ETG samples is shown by the solid orange curve (median), with the error bars denoting the [$16\%$, $84\%$] distribution of $\gamma^{\prime}$. They have stellar masses within the central $30\,\mathrm{kpc}$ $\mathrm{log}\,(M_{\ast}/\mathrm{M_{\astrosun}})\in\,$[$10.7$, $11.9$] in all selected redshifts from $z=0$ to $z=3$. The horizontal grey dashed line stands for the exact isothermal slope $\gamma^{\prime}=2$.}
\label{fig:tracer_TOT}
\end{figure}

We first present the evolutionary trend of $\gamma^{\prime}$ of the MPB sample and the statistical ETG sample in the full stellar mass range $\mathrm{log}\,(M_{\ast}/\mathrm{M_{\astrosun}})\in[10.7, 11.9]$. The evolutionary trends of the total power-law density slope $\gamma^{\prime}$ for IllustrisTNG ETGs are shown in Fig.~\ref{fig:tracer_TOT}. 

\begin{figure*}
\includegraphics[width=2\columnwidth]{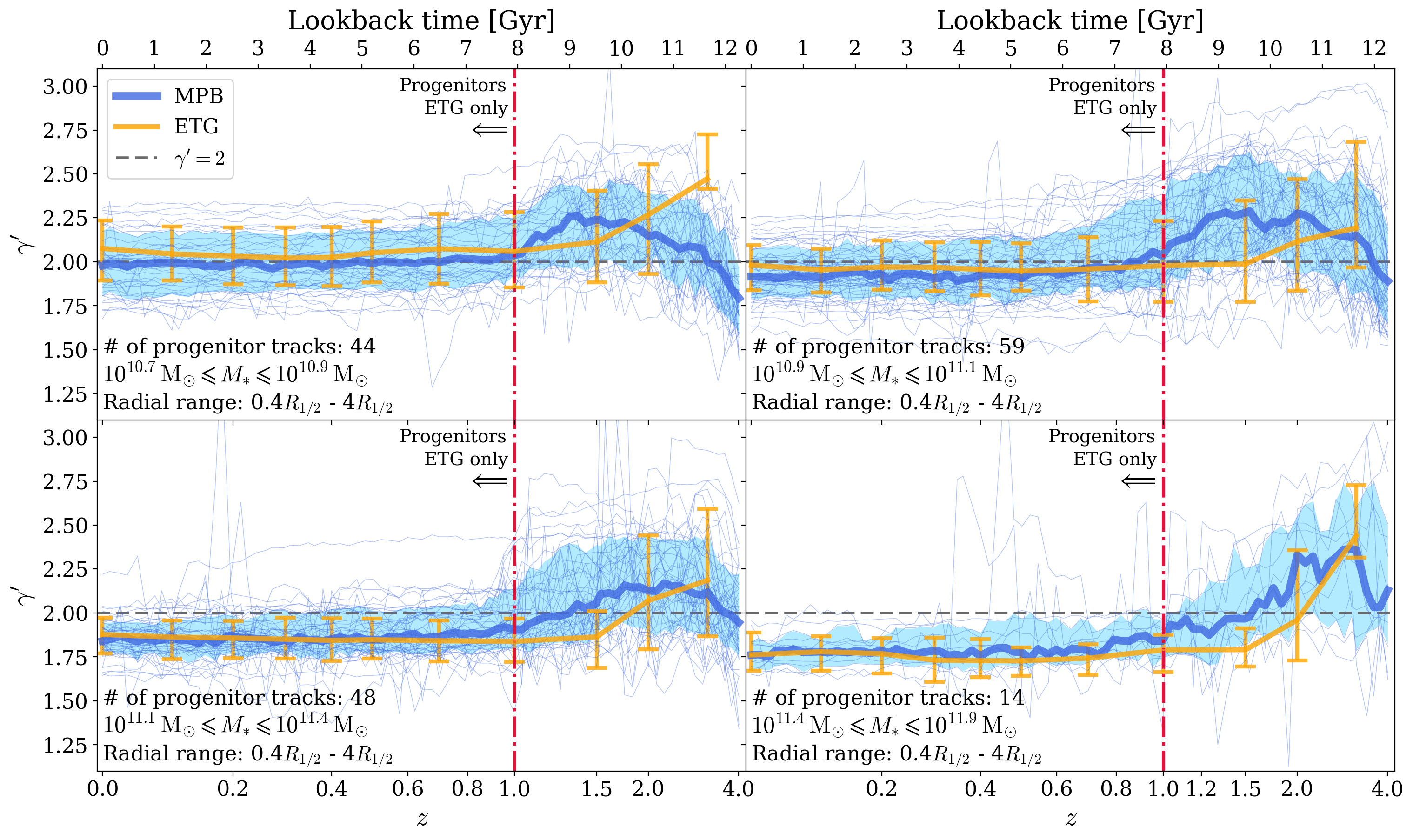}
\caption{The evolutionary trend of the total power-law slope $\gamma^{\prime}$ with redshift, divided in 4 stellar mass (enclosed within central $30\,\mathrm{kpc}$) bins, i.e. $\mathrm{log}\,(M_{\ast}/\mathrm{M_{\astrosun}})\in\,$[$10.7$, $10.9$], [$10.9$, $11.1$], [$11.1$, $11.4$], and [$11.4$, $11.9$]. The MPB tracing sample is shown in blue, divided into 4 bins based on their $z=0$ stellar mass. The thick solid curves denote the median and the shaded regions denote the [$16\%$, $84\%$] distribution of $\gamma^{\prime}$. The individual evolutionary tracks of $\gamma^{\prime}$ in the progenitors are also shown by the thin blue solid lines traced from $z=0$ to $z=4$. Their selection forces them to have already become ETG morphologically by $z=1$, shown by the red dashed dotted lines. The statistical ETG sample from $z=0$ to $z=3$ is shown by the solid orange curves (median), with the error bars denoting the [$16\%$, $84\%$] distribution of $\gamma^{\prime}$. The horizontal grey dashed line stands for the exact isothermal slope $\gamma^{\prime}=2$.}
\label{fig:tracer_bins}
\end{figure*}

The MPB tracing sample is shown by the blue solid curve and shaded region, which represents the median and [$16\%$, $84\%$] distribution, respectively. As can be seen, roughly before $z=2$ the total density slope $\gamma^{\prime}$ steepens with time as a consequence of secular star formation and wet mergers. Between $z\approx1$ and $z\approx2$, $\gamma^{\prime}$ gradually becomes shallower, as a consequence of AGN feedback (see Section~\ref{sec: 4.2} for details). Below $z=1$, the total profile passively evolves with little change, staying on average close to an isothermal distribution. We note that the MPB $\gamma^{\prime}$ evolution has large intrinsic scatter, and it is consistent with an isothermal total density profile within the $1\,\sigma$ scatter from redshift 4 and 0. 

The statistically selected samples of ETGs at various redshifts from $z=0$ to $z=3$ are shown by the orange solid curve (median), with the error bars denoting the [$16\%$, $84\%$] distribution. The total density profile demonstrates a different evolutionary trend compared to the MPB tracing sample. Above $z=1$, the median value of the total power-law density slope starts out significantly steeper than isothermal, and then decreases with time. Below $z = 1$, the statistical ETG sample slope shows almost no evolution. However, a slight but noticeable increase in the mean $\gamma^{\prime}$ towards $z=0$ is present (below $z=0.5$), which is interestingly in line with observational evidence of the total density profile evolution towards low redshift.~\citep{2011ApJ...727...96R,2012ApJ...757...82B,2013ApJ...777...98S,2014MNRAS.440.2013D,2017ApJ...840...34S,2018MNRAS.480..431L}. 
Although both the progenitor tracing and the statistical samples show little evolution of $\gamma^{\prime}$ below $z=1$, their median evolution trends with time are notably different. In Section~\ref{sec: 3.3} and Section~\ref{sec: 3.4}, we consider the factors that could account for such differences. We remind the reader that the scatter of $\gamma^{\prime}$ in both samples is large, rendering both samples' total density profile evolutionary trend consistent with a constant isothermal density profile, and statistical uncertainties also play a role in the difference of the two evolution trends. 

\subsection{The mass dependence of the slope evolution}
\label{sec: 3.3}

In order to understand the mass dependence of the slope evolution, we show the $\gamma^{\prime}$ evolutionary trends in the four different stellar bins in Fig.~\ref{fig:tracer_bins}. The $\gamma^{\prime}$ evolution of individual MPB tracks is plotted with the thin blue curves. Their median and [$16\%$, $84\%$] distribution are shown by the thick blue curve and the shaded region, respectively. The solid orange curve denotes the median of the statistical ETG sample $\gamma^{\prime}$, with the error bars standing for their [$16\%$, $84\%$] distribution. We note that the scatter of $\gamma^{\prime}$ in both samples among the four stellar mass bins is large, especially in the two lower mass bins and at redshifts above $z\approx1$.

As can be seen in all four mass bins, both statistical (yellow) and MPB (blue) distributions show constant $\gamma^{\prime}$ evolution since $z=1.0$. The average $\gamma^{\prime}$ within this redshift range strongly depends on galaxy mass: on average lower-mass galaxies have steeper total density profiles than their more massive counterparts (also see Fig. 4 in \citealt{2018arXiv181106545W}). 

The MPB tracing tracks (blue) for more massive present-day galaxies have $\gamma^{\prime}$ peaking at higher redshifts, indicating they left the in-situ star formation/wet merger phase earlier than their lower-mass counterparts. Note that as high-$z$ lower-mass galaxies in the statistical sample are among the MPBs of present-day galaxies in the higher-mass bins, the mass-dependence of the $\gamma^{\prime}$-peak also explains the differences seen at higher redshifts between the MPB tracing samples (blue) and the statistical samples (yellow), which is most established in the lowest mass bin. For the same reason, the two samples also show different redshift dependencies at $z>1$ in Fig.~\ref{fig:tracer_TOT}, as lower-mass galaxies always dominate any given galaxy sample. Deviations from an isothermal total density profile at the $\sim1\sigma$ scatter level are present for both samples in the redshift range $z\gtrsim 1.5$, as well as in $z\lesssim 0.8$ for both samples in the two higher stellar mass bins. 

\subsection{The dependence of the averaged $\gamma^{\prime}$ on ETG transition redshift}
\label{sec: 3.4}

As described in Section~\ref{sec: 3.1}, we require our MPB tracing sample to be composed of galaxies whose progenitors as early as $z=1$ must also be ETGs. This results in zero-redshift $\langle \gamma^{\prime} \rangle = 1.904 \pm 0.012$ and $\sigma_{\gamma^{\prime}} = 0.149$ for our MPB tracing sample, which is slightly shallower than the isothermal profiles in observational samples~\citep{2006ApJ...649..599K,2010ApJ...724..511A,2011MNRAS.415.2215B,2011ApJ...727...96R,2013ApJ...777...98S,2016MNRAS.460.1382S,2017MNRAS.467.1397P,2018MNRAS.476.4543B}. 

\begin{figure}
\includegraphics[width=\columnwidth]{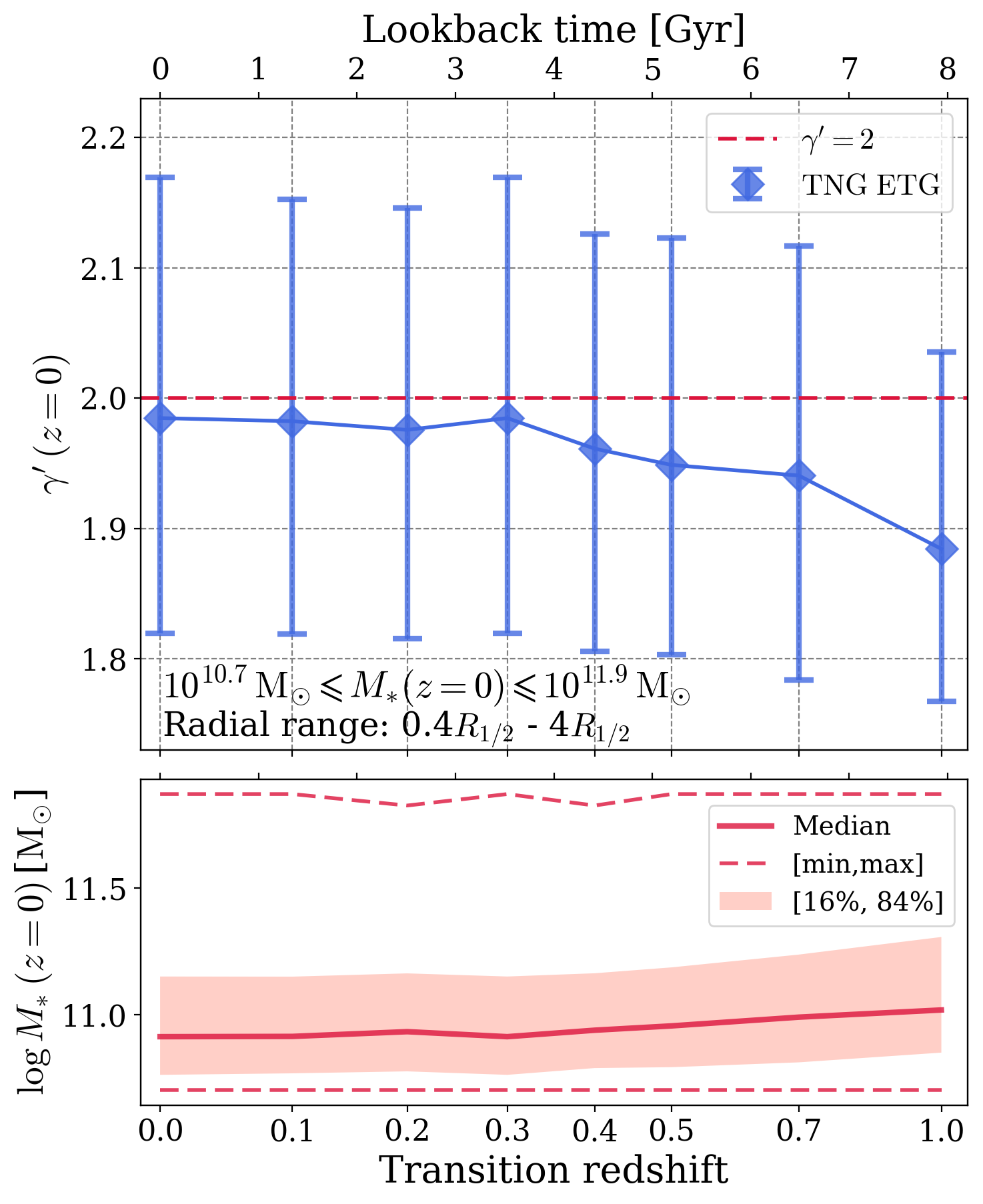}
\caption{Upper panel: zero-redshift $\gamma^{\prime}$ as a function of the redshift at which the main progenitors also have to be ETGs. The diamonds indicate the median $\gamma^{\prime}$ and the error bars indicate the [$16\%$, $84\%$] distribution of $\gamma^{\prime}$. The horizontal red dashed line stands for the isothermal slope $\gamma^{\prime}=2$. Bottom panel: the median (solid line), [$16\%$, $84\%$] distribution (shaded region), and the minimum/maximum values (dashed lines) of the $z=0$ stellar mass for the selected samples that quenched at different redshifts. }
\label{fig:quench}
\end{figure}

In order to understand the dependence of the averaged $\gamma^{\prime}$ on ETG transition redshift, we trace all ETGs in the $z=0$ statistical sample along their main progenitor branch, and require the progenitors at different investigated redshifts, i.e. $z = 0.1, 0.2, 0.3, 0.4, 0.5, 0.7, 1.0$ to be also ETGs. However, we do not restrict galaxy types for the progenitors prior to each one of these redshifts. We then take the selected samples at different redshifts and trace them forward to $z=0$ to obtain the zero-redshift density slope distribution. The result is presented in Fig.~\ref{fig:quench}. As can be seen, the later in time (lower-$z$) that we require the progenitors to become ETGs, the steeper the median zero-redshift $\gamma^{\prime}$ becomes. The small bump of the $\gamma^{\prime}(z = 0)$ evolution at $z=0.3$ arises from the inclusion of slightly more lower-mass galaxies in the sample that made their transition to ETGs at $z=0.3$ (lower panel in Fig.~\ref{fig:quench}). Thus, the zero-redshift total power-law density slope distribution is affected by the galaxy sample's transition time. However, this trend is rather mild given that the large intrinsic scatter in the distribution of $\gamma^{\prime}(z=0)$ is consistent with little evolution with the galaxy type-transition time shown by the [$16\%$, $84\%$] interval in Fig.~\ref{fig:quench}. To analyze the impact of sample transition time on zero-redshift median value of $\gamma^{\prime}$, we consider the consequence of two effects: (1) lower-mass ETGs have steeper slopes than their more massive counterparts (see Fig.~\ref{fig:tracer_bins}); (2) \citet{2018MNRAS.474.3976G} found for the TNG galaxies that the transition to ETGs (quenched galaxies) moves towards lower-mass galaxies with cosmic time (see their Fig. 4), which is consistent with observations (e.g. \citealt{2008ApJ...688..770F,2015ApJ...813...23V,2015MNRAS.448..237W,2016MNRAS.462.2559B,2017ApJ...838...19W}). The inclusion of more and more lower-mass galaxies in the ETG sample results in steepened slopes towards $z=0$. This may also explain the lower-$z$ steepening trend of the statistical sample in Fig.~\ref{fig:tracer_TOT}.

\subsection{Example $\gamma^{\prime}$ evolution with mock galaxy images}
\label{sec:3.5}

\begin{figure*}
\includegraphics[width=1.8\columnwidth]{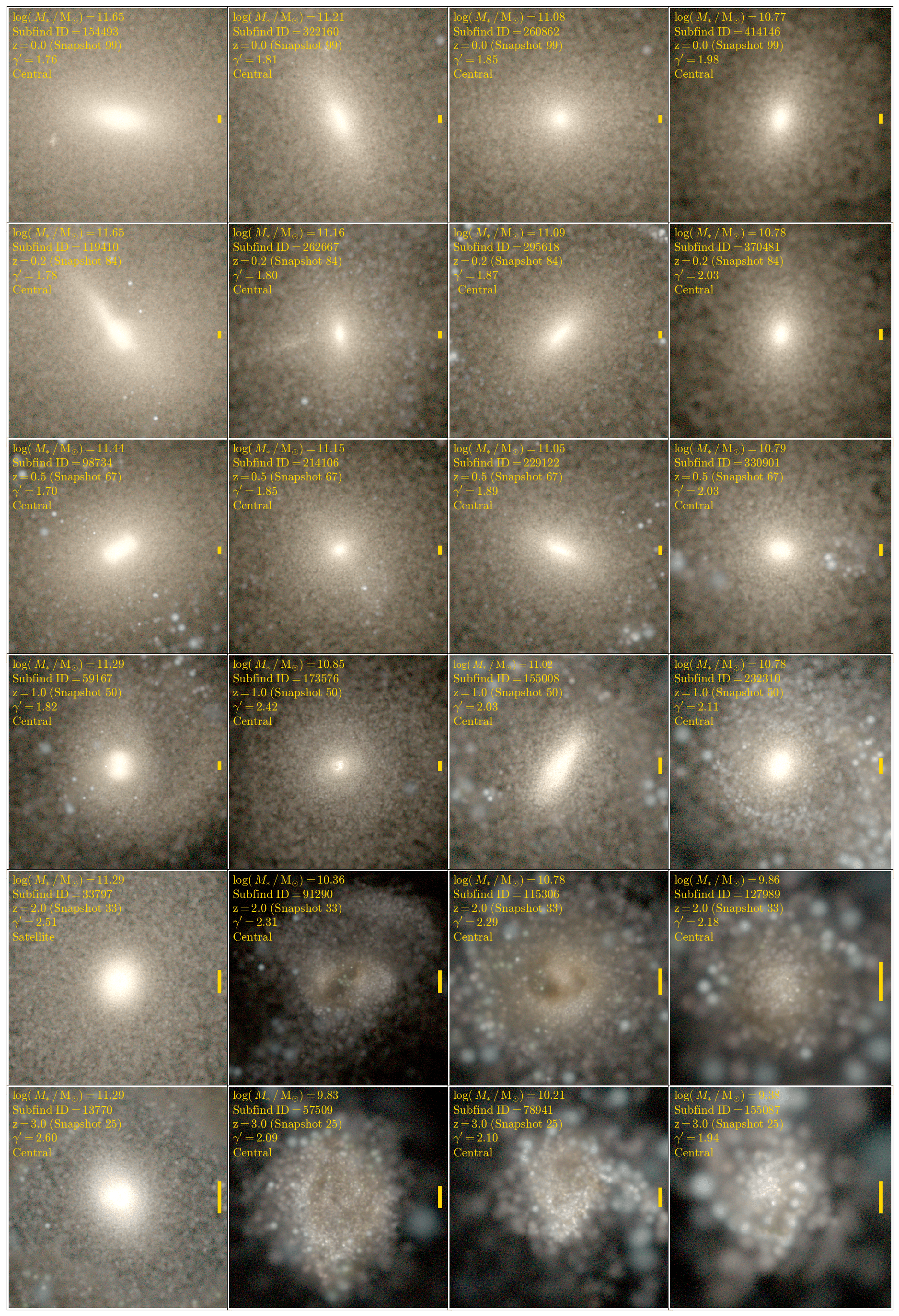}
\caption{Illustration of individual MPB evolution (bottom to top) for 4 example ETGs, one selected from each of the 4 stellar mass bins. The images are the rest frame SDSS $g,r,i$-band composite mock images of IllustrisTNG ETGs projected along their face-on direction. The images also consider dust attenuation by adopting a resolved dust radiative transfer model with dust-to-metal ratio of $0.3$~\citep{2019arXiv190407238V}. A yellow scale bar depicting $1\,\mathrm{kpc}$ is plotted to the right side of each image. As it can be seen, the more massive galaxies transform to ETGs earlier, are less dusty at high redshift, and more diffuse toward lower redshift; vice versa for less massive galaxies.}
\label{fig:combined}
\end{figure*}

To visualize galaxy evolution which encodes the evolution of their total power-law density slopes, we select 4 example galaxies from the MPB sample, one in each of the 4 stellar mass bins, and present their mock images at redshifts $z= 0, 0.2, 0.5, 1, 2, 3$ in Fig.~\ref{fig:combined}. These are the rest frame SDSS $g,r,i$-band composite images with the galaxies projected in their face-on direction (defined as the direction of the mass-weighted stellar angular momentum of the galaxy). The images also consider dust attenuation by adopting a resolved dust radiative transfer model with dust-to-metal ratio of $0.3$~\citep{2019arXiv190407238V}. Each image is cropped to a side length $l=\mathrm{min}(6\,R_{1/2}, 30\,\mathrm{kpc})$, where $R_{1/2}$ is the 3D stellar half mass radius of the galaxy, and $30\,\mathrm{kpc}$ is in physical units. A yellow scale bar depicting physical $1\,\mathrm{kpc}$ is shown on the right side of each image. Each galaxy evolves from the bottom ($z=3$) to the top ($z=0$) in the figure. The stellar mass $M_{\ast}$, Subfind ID, redshift $z$ (snapshot number), total power-law density slope $\gamma^{\prime}$, and the group identity of being a central or satellite galaxy are labeled for every main progenitor we present in the figure.

As shown by the mock images and the labeled properties, the largest ETG at $z=0$ has progenitors with shallower total density profile compared to the smallest ETG at $z=0$. The smaller ETGs also transform from late-type to early-type later than the larger ETGs, indicative of progenitor bias for selecting statistical ETG samples at lower redshift. The 4 example tracks also capture a few ongoing or recently occurred merger events in the limited snapshots (e.g. Snapshot 84, Subfind ID 119410). It is seen that heavier (lighter) ETG end products at $z=0$ are more visually diffuse (compact), which is consistent with the $\gamma^{\prime}-\mathrm{log}\,M_{\ast}$ correlation at $z=0$ found in Paper I (Fig. 4), demonstrating that the larger stellar mass ETGs at $z=0$ have shallower $\gamma^{\prime}$. An obvious color transition from blue to red for the galaxies also reflects the type transition from late-type to early-type of the main progenitors.

\section{Effects of galaxy mergers and AGN feedback on $\gamma^{\prime}$}
\label{sec: 4}

In this section, we study the effects of galaxy mergers and AGN feedback on the evolution of the total density profile for the MPB sample. We also propose a theoretical ETG formation scenario within the IllustrisTNG Simulation context that traces the evolution of $\gamma^{\prime}$ to mark three distinct formation phases. 
\subsection{Galaxy mergers}
\label{sec: 4.1}

\begin{figure}
\includegraphics[width=\columnwidth]{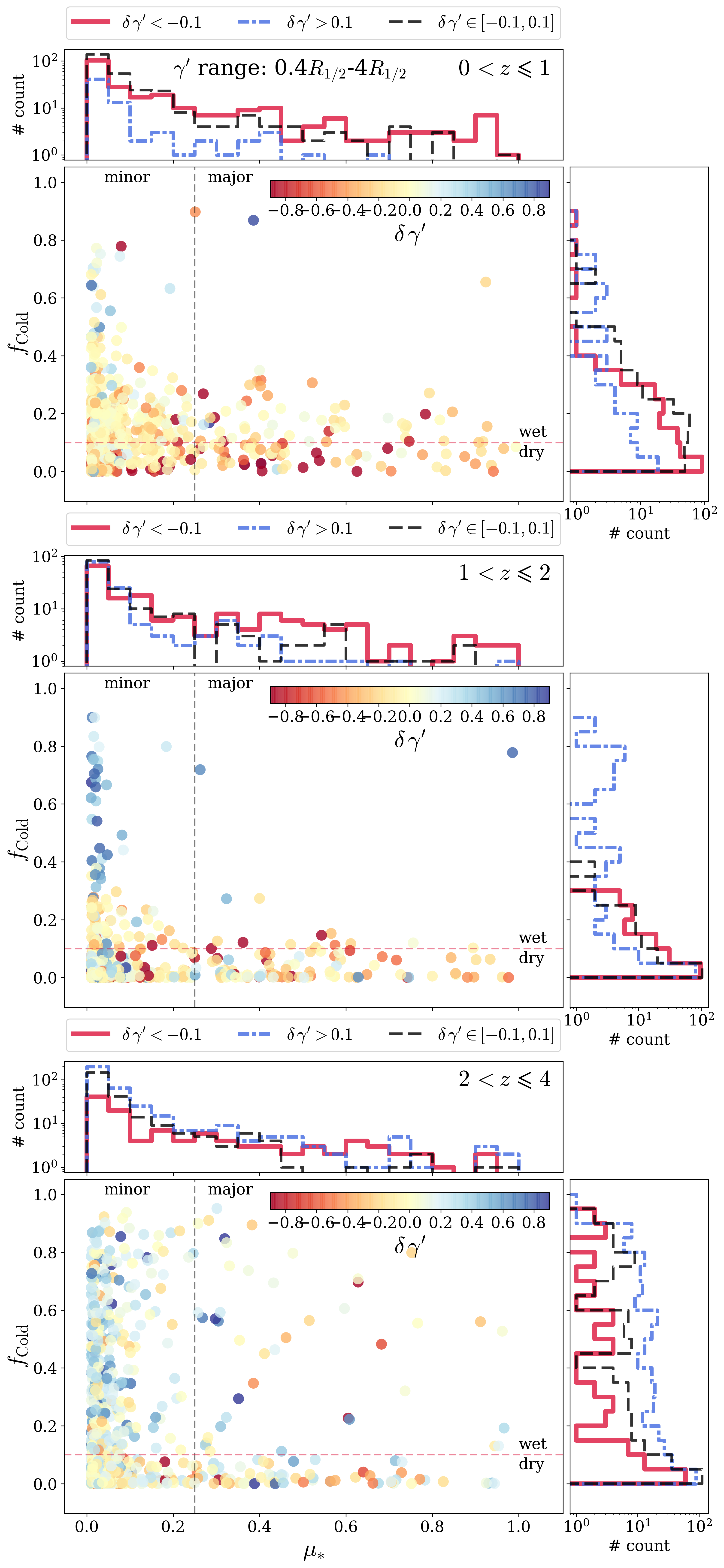}
\caption{The cold gas fraction $f_{\mathrm{Cold}}$ versus the merger mass ratio $\mu_{\ast}$ for all mergers that occurred to the 165 main progenitors traced from $z=0$ to $z=4$ are shown by the scattered dots in the figure. The dots are color coded by the change in the density profile $\delta\gamma^{\prime}$ before and after a given galaxy merger. We divide all merger events for all main progenitors into three redshift bins, i.e. $z\in [0, 1], [1,2],\ \mathrm{and}\ [2,4]$, which are shown in the three subplots from top to bottom, respectively. The number distribution of mergers according to $\mu_{\ast}$ and $f_{\mathrm{Cold}}$ are shown by the histograms projected in the $X$ and $Y$ directions in each subplot. Each set of projected merger number distribution histograms is divided into three broad bins with $\delta\gamma^{\prime} < -0.1$ (red solid histogram), $\delta\gamma^{\prime}\in[-0.1, 0.1]$ (black dashed histogram), and $\delta\gamma^{\prime}>0.1$ (blue solid histogram), corresponding to the total density profile becoming `shallower', `invariant', and `steeper' due to galaxy mergers.}
\label{fig:merger}
\end{figure}

Merger processes are often classified as wet (dissipative) or dry (non-dissipative). Dissipation is widely present in galaxies with significant gas fraction (typically late-type). Gas eventually loses pressure support and falls in, leading to enhanced star formation activity in the inner regions of the galaxy and the total density profile becomes steeper. In contrast, dry mergers gradually build up the outskirts of a galaxy (minor) and smooth out the matter distribution (major) due to lack of dissipation, making the total density profile shallower or nearly invariant after the merger.

Merger trees for IllustrisTNG galaxies examined here are constructed with the \textsc{sublink} algorithm \citep{2015MNRAS.449...49R} as described in Section~\ref{sec: 2.2}. Although observationally mergers are classified into gas-rich and gas-poor according to the pair U-B color following \citet{2008ApJ...681..232L}, we adopt the definition for `wet' and `dry' mergers from \citet{2009MNRAS.397..506K} which is a more physical definition involving galaxy intrinsic properties available in the output from the simulation. We define mergers with cold gas fraction $f_{\mathrm{Cold}}$ satisfying:
\begin{equation}
f_{\mathrm{Cold}} = \frac{M_{\mathrm{Cold}}}{M_{\mathrm{Cold}} + M_{\ast}} \leqslant 0.1
\end{equation}
to be dry mergers, and otherwise wet mergers. Here, $M_{\mathrm{Cold}}$ is defined as the combined mass of all gas cells assigned to the main progenitor and the infall progenitor that have temperature $T<10^{4}\,\mathrm{K}$. $M_{\ast}$ here is also the combined stellar mass of the main and infall progenitors, each enclosed within twice their 3D stellar half mass radii.

To analyze the impact of merger events on the total power-law density slope, we derive the change in $\gamma^{\prime}$ for each merger event:
\begin{equation}
\delta\gamma^{\prime} = \gamma^{\prime}_{\mathrm{Des}} - \gamma^{\prime}_{\mathrm{Main}}\,,  
\end{equation}
where $\gamma^{\prime}_{\mathrm{Des}}$ and $\gamma^{\prime}_{\mathrm{Main}}$ are the total power-law density slopes of the descendant galaxy and the main progenitor, respectively. From the definition of $\delta\gamma^{\prime}$, a positive (negative) $\delta\gamma^{\prime}$ indicates that the descendant has steeper (shallower) density profile than its main progenitor due to merger impact. The descendant is usually identified $1-2$ snapshots after the merger, and the main progenitor is usually identified a few snapshots before the merger where the infall progenitor's stellar mass reached its maximum. Thus the cadence in which $\delta\gamma^{\prime}$ is measured is $2-5$ snapshots ($\sim 371\,\mathrm{Myr}$ at $z=0.2$, $\sim 348\,\mathrm{Myr}$ at $z=1$, and $\sim 294\,\mathrm{Myr}$ at $z=3$) in most cases, which is about the order of the typical dynamical timescale ($\sim 100\,\mathrm{Myr}$) of the main progenitors.

The cold gas fraction $f_{\mathrm{Cold}}$ versus merger mass ratio $\mu_{\ast}$ color coded by the change in slope $\delta\gamma^{\prime}$ for all mergers of our 165 ETGs traced from $z=0$ to $z=4$ is shown by the scattered dots in Fig.~\ref{fig:merger}. According to the overall evolutionary trend of the average total power-law density profile for the MPB sample in Fig.~\ref{fig:tracer_TOT}, $\gamma^{\prime}$ steadily rises from $z=4$ to $z\approx2$, drops steadily from $z\approx2$ to $z\approx1$, and becomes nearly constant below $z\approx1$. Thus, we divide all merger events for all main progenitors into 3 redshift bins, i.e. $z\in [0, 1], [1,2],$ and $[2,4]$, which are shown in the 3 subplots from top to bottom in Fig.~\ref{fig:merger} correspondingly. The number distribution of mergers according to merger mass ratio $\mu_{\ast}$ and merger cold gas fraction $f_{\mathrm{Cold}}$ are shown by the histograms projected in the X and Y directions in each subplot. Each set of projected merger number distribution histograms is divided into 3 bins with $\delta\gamma^{\prime} < -0.1$ (red solid histogram), $\delta\gamma^{\prime}\in[-0.1, 0.1]$ (black dashed histogram), and $\delta\gamma^{\prime}>0.1$ (blue solid histogram), corresponding to the total density profile becoming `shallower', `invariant', and `steeper' as a result of galaxy mergers.

As shown in the figure, the distributions of $\mu_{\ast}$, $f_{\mathrm{Cold}}$, and $\delta\gamma^{\prime}$ are significantly different in the 3 redshift ranges. In the redshift range $z\in[2, 4]$ (bottom panel), minor mergers dominate the merger events, and a large proportion of these mergers are wet, resembling the gas-rich phase of galaxy assembly above $z\approx2$. From the color coding of the scattered dots as well as the $\delta\gamma^{\prime}-f_{\mathrm{Cold}}$ histogram, the wet mergers induce a steepening effect of the total density profile. The effect on $\delta\gamma^{\prime}$ with different $\mu_{\ast}$ is insignificant during $z\in[2,4]$. In the redshift range $z\in[1,2]$ (middle panel), minor and major mergers are almost free of wet mergers, although a small proportion of minor mergers remain wet, which resembles a transition period where late-type galaxies gradually turn into early-type galaxies. The color coded scattered dots and the $\delta\gamma^{\prime}-f_{\mathrm{Cold}}$ histogram clearly show that mergers with smaller (larger) $f_{\mathrm{Cold}}$ result in shallower (steeper) density profiles as a result of the mergers. Since during $z\in[1,2]$ the major mergers are dominated by dry mergers, larger (smaller) $\mu_{\ast}$ results in shallower (steeper) slope after the mergers. In the redshift range $z\in[0,1]$, the majority of mergers have $f_{\mathrm{Cold}} \lesssim 0.4$, and the proportion of wet minor mergers also decreases. Since our sample selection enforces the entire MPB sample to be transformed into ETGs by $z=1$, the evolution of the progenitors during $z\in[0,1]$ resembles the passive evolution of ETGs below $z=1$ that is dominated by gas-poor mergers (see also \citealt{2007ApJ...658..710N}). The color coded scattered dots and the two histograms reveal that gas-poor mergers (i.e. $f_{\mathrm{Cold}}\lesssim 0.4$) with larger merger mass ratio result in a shallower or unchanged  $\gamma^{\prime}$. Appendix \ref{sec: A} shows further analyses of the statistical significance between galaxy mergers and the change of the total density profile. 

Since galaxies assemble about $70\%$ of their ex-situ stellar mass through major and minor mergers~\citep[$\mu\geqslant 1/10$]{2016MNRAS.458.2371R}, the dominance of dry mergers which correlates with $\delta\gamma^{\prime}<0$ (especially in $z\leqslant 2$ ) in this merger mass ratio range is expected to make the total density profile shallower with time. This is consistent with the positive $\gamma^{\prime}-f_{\mathrm{in-situ}}$ correlation found for $z=0$ IllustrisTNG ETGs in Paper I, and the stellar assembly of the galaxy could be heavily dominated by ex-situ stellar populations accreted in mergers that make up $\gtrsim 80\%$ of the total stellar mass in the most massive IllustrisTNG ETGs (see Fig. 10 in \citealt{2018arXiv181106545W}). Since most of the major mergers and a part of the minor mergers are dry, the $\gamma^{\prime}$ in galaxies with lower $f_{\mathrm{in-situ}}$ become shallower compared to galaxies with higher $f_{\mathrm{in-situ}}$, creating the positive $\gamma^{\prime}-f_{\mathrm{in-situ}}$ trend.

Overall, the impact of mergers on the density profile with different stellar mass ratios and cold gas fractions for IllustrisTNG ETGs is consistent with predictions from previous works~\citep{2009ApJ...690..802J,2012ApJ...754..115J,2013ApJ...766...71R,2014ApJ...786...89S}. The increasing importance of dry mergers at $z\lesssim 1$ driving ETG passive evolution in our results is also consistent with the merger-induced spin-down of ETGs constrained through observations~(e.g. \citealt{2018ApJ...862..125N}). However, we point out that due to the limited spatial and mass resolution in the TNG100 simulation (where most of the infall progenitors are not well-resolved), the physical picture of a galaxy merger impact on the total density profile that we have found for the IllustrisTNG ETGs still requires finer characterization with high-resolution zoom-in simulations. This is especially important for a better understanding of the impact of `dry' mergers and the correlation between $\mu_{\ast}$ and $f_{\mathrm{Cold}}$ which calls for well-resolved properties of the infall progenitors in galaxy mergers. 

\subsection{AGN feedback}
\label{sec: 4.2}

\begin{figure}
\includegraphics[width=\columnwidth]{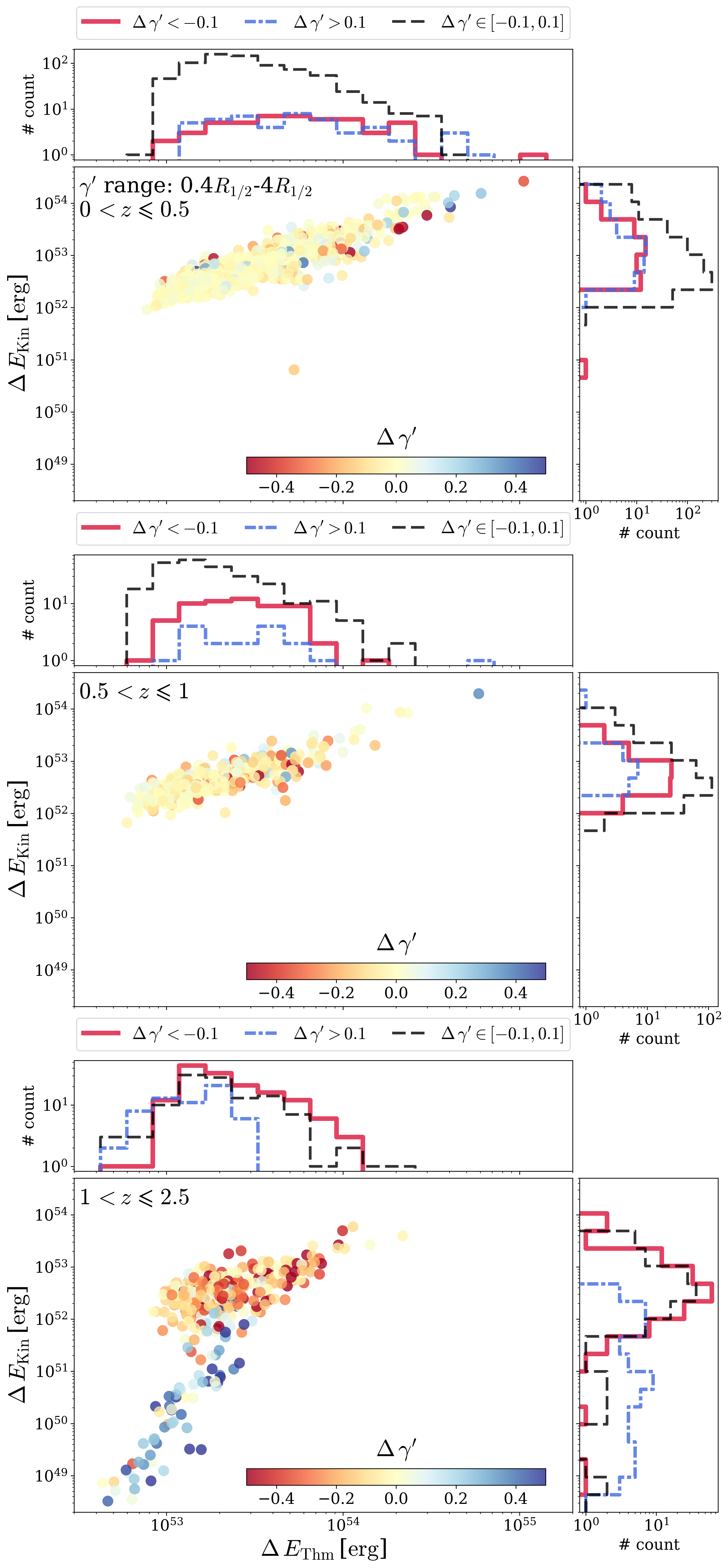}
\caption{This figure shows the impact of AGN feedback energy on the change of $\gamma^{\prime}$. Each scattered dot corresponds to the $\Delta E_{\mathrm{Kin}}$ and $\Delta E_{\mathrm{Thm}}$ which are color coded by $\Delta\gamma^{\prime}$, all measured within a time scale of $\sim 1$ Gyr. We divide the time evolution into three epochs, i.e. $z \in [0, 0.5], [0.5, 1.0],,\ \mathrm{and}\ [1.0, 2.5]$, which are shown in the three subplots from top to bottom in the figure. The number distribution of AGN feedback energy in the kinetic and thermal modes are shown by the histograms projected in the $X$ and $Y$ directions in each subplot. Each set of projected feedback energy distribution histograms is divided into three broad bins with $\Delta\gamma^{\prime} < -0.1$ (red solid histogram), $\Delta\gamma^{\prime}\in[-0.1, 0.1]$ (black dashed histogram), and $\Delta\gamma^{\prime}>0.1$ (blue solid histogram), corresponding to the total density profile becoming `shallower', `invariant', and `steeper' due to AGN feedback in the given time intervals.}
\label{fig:bh_tracks}
\end{figure}

AGN feedback plays an important role in quenching star formation and transforming galaxies from late-type to early-type, especially in the case of the Bondi accretion model of black hole growth adopted in IllustrisTNG~\citep{2017MNRAS.465...32B,2017MNRAS.465.3291W}. One piece of observational evidence for supermassive black holes regulating galaxy evolution is the $M_{\mathrm{BH}}-\sigma_{\mathrm{v}}$ scaling relation in ETGs~\citep{2001ApJ...552L..13C,2003ApJ...596..903P,2006ApJ...641...90R,2008ApJ...680..143G,2013ARA&A..51..511K,2013ApJ...764..184M}. Apart from this observational scaling relation, AGN feedback is found essential to reproduce a realistic redshift evolution of the total density profiles in numerical simulations, e.g. the Illustris Simulations~\citep{2017MNRAS.469.1824X}, the Magneticum Pathfinder~\citep{2017MNRAS.464.3742R}, and the Horizon-AGN Simulations~\citep{2017MNRAS.472.2153P,2019MNRAS.483.4615P}.

Recent developments in cosmological simulations including AGN feedback have explored a wide range of AGN feedback models. The AGN feedback model of the Horizon-AGN Simulations~\citep{2014MNRAS.444.1453D} was described in \citet{2012MNRAS.420.2662D}. Unlike the \citet{2007MNRAS.380..877S} model adopted for Illustris, where the low accretion rate AGN feedback deposits thermal energy in the form of radio bubbles that mimic the cocoon-like radio lobes observed in clusters, \citet{2012MNRAS.420.2662D} models the bi-polar outburst from the low-accretion phase of AGN feedback directly by depositing momentum and energy in a cylindrical-shaped vicinity of the supermassive black hole. Recent observational evidence~\citep{2016Natur.533..504C} has revealed bi-symmetric structures in the central regions of quiescent galaxies in the MaNGA Survey~\citep{2015ApJ...798....7B}, in support of radio mode AGN-driven bi-polar jets. However, they also infer the presence of centrally driven winds, which could mechanically suppress star formation in quiescent galaxies. The low accretion rate kinetic mode AGN feedback in the IllustrisTNG black hole model~\citep{2017MNRAS.465.3291W,2018MNRAS.479.4056W} is a direct effort to model the centrally-driven mechanical winds, and its major difference with the radio modes of \citet{2007MNRAS.380..877S} and \citet{2012MNRAS.420.2662D} is that the kinetic mode of the IllustrisTNG model ejects only momentum and pure kinetic energy into its surrounding gas cells while no immediate thermal energy is deposited. This is in line with the results from recent high resolution hydrodynamical simulations of isolated AGN in a single elliptical galaxy, where the AGN is found to be mostly operating in the quiescent phase, and this highlights the importance of kinetic winds and jets in producing a realistic galaxy that matches observations~\citep{2018ApJ...857..121Y, 2019arXiv190107570Y}. The high accretion rate thermal mode AGN feedback in the IllustrisTNG model follows the common practice of previous works (also know as the quasar mode, \citealt{2005MNRAS.361..776S,2007MNRAS.380..877S,2009MNRAS.398...53B,2012MNRAS.420.2662D,2013MNRAS.436.3031V}) which injects pure thermal energy to the local interstellar medium of the central supermassive black hole. 

Based on the variety of above-mentioned AGN feedback implementations, previous work has highlighted the important link between AGN feedback and galaxy properties including the total density profile. \citet{2008A&A...479..123P} found that bi-polar outbursts from AGN feedback can drive oscillation of gas cores in the center of galaxies that flattens the density profiles of dark matter and stars within a timescale of $4\sim5$ Gyrs. \citet{2010MNRAS.405.2161D} found that AGN feedback leads to shallower-than-isothermal cluster inner density profiles and reduces baryon fractions. \citet{2012MNRAS.420.2859M} proposes a new mechanism involving dynamical friction heating and gaseous ejections from AGN feedback that creates cored stellar profiles and forms near-isothermal total density profiles in luminous elliptical galaxies, similar to the findings of \citet{2008A&A...479..123P}. \citet{2012MNRAS.422.3081M} further points out that the slow expulsion of gas in the quiescent (low-accretion rate) phase of AGN activity can also lead to adiabatic expansion that further flattens the total density profile. Furthermore, AGN feedback quenches in-situ star formation, enhances the accreted stellar populations from mergers at larger radii, and removes baryonic matter from galaxy central regions. This process typically consists of dynamically inducing radial stellar orbits~\citep{2013MNRAS.433.3297D,2015ApJ...804L..40G} and injecting momentum into the surrounding interstellar medium which creates AGN-driven outflows \citep{2012MNRAS.425..605F,2012MNRAS.424..190G,2016MNRAS.458..816H,2017ApJ...835...15C,2017MNRAS.465.3291W}. 

Apart from AGN feedback, feedback from stellar winds may also have an impact on the total density profile. In the absence of AGN feedback, stronger stellar feedback can lead to enhanced in-situ-formed stellar populations~\citep{2013MNRAS.436.2929H} which correlates with larger $\gamma^{\prime}$~\citep{2013ApJ...766...71R}, hence leading to steeper total density profiles~\citep{2017MNRAS.464.3742R}. Nevertheless, we have shown using the various TNG model variations in Section 4 of Paper I that, with the presence of AGN feedback, stronger stellar winds lead to shallower total density profiles for IllustrisTNG ETGs, opposite to the effects of stellar winds without AGN feedback. Another important finding of Paper I is that the impact of stellar wind feedback on changing $\gamma^{\prime}$ is subdominant compared to AGN feedback. Hence, we will focus our discussion of the relation between feedback  and $\gamma^{\prime}$ on AGN feedback as it is the most prominent.

To study the AGN feedback impact on the density profiles of IllustrisTNG ETGs, we investigate the feedback energy injected by the central massive black hole of the main progenitor in two channels, the high accretion rate thermal mode and the low accretion rate kinetic mode. In particular, to obtain the effect of AGN feedback energy on the total density profile, we divide the $z=0$ to $z=4$ evolution tracks into 9 segments (time intervals), roughly 1Gyr for each time interval, and plot the energy injected by the central supermassive black hole in the kinetic and thermal AGN feedback modes versus the change of $\gamma^{\prime}$ in these time intervals in Fig.~\ref{fig:bh_tracks}. The time interval boundaries are set at 10 redshifts 0.00, 0.07, 0.15, 0.24, 0.35, 0.52, 0.68, 1.00, 1.50, 2.44 corresponding to lookback times 0.00, 1.01, 1.98, 2.97, 3.97, 5.37, 6.35, 7.92, 9.51, 11.18 Gyrs. The 9 time intervals are then set between these 10 boundaries. Similar to Fig.~\ref{fig:merger}, we divide the time intervals into 3 redshift ranges, i.e. $z \in [0, 0.5], [0.5, 1.0],,\ \mathrm{and}\ [1.0, 2.5]$, which are shown in the three subplots from top to bottom in Fig.~\ref{fig:bh_tracks}. The reason for not following the third redshift bin up to $z=4$ is due to the presence of `halo-switch' jumps in every snapshot for $z\in[2.5, 4]$, making it very difficult to follow the AGN feedback energy of the full MPB sample in this redshift range.
 
For each main progenitor, the AGN feedback energies injected in the kinetic and thermal modes for a given time interval are denoted by $\Delta E_{\mathrm{Kin}}$ and $\Delta E_{\mathrm{Thm}}$. The change of the total power-law density profile $\Delta\gamma^{\prime}$ in that period is evaluated as the change in $\gamma^{\prime}$ at the beginning and the end snapshot of that time interval. Each scattered dot in Fig.~\ref{fig:bh_tracks} corresponds to the $\Delta E_{\mathrm{Kin}}$ and $\Delta E_{\mathrm{Thm}}$ which are color coded by $\Delta\gamma^{\prime}$ in a given time interval. The number distribution of AGN feedback energy in the kinetic and thermal modes are shown by the histograms projected in the X and Y directions in each subplot. Each set of projected feedback energy distribution histograms is divided into 3 bins with $\Delta\gamma^{\prime} < -0.1$ (red solid histogram), $\Delta\gamma^{\prime}\in[-0.1, 0.1]$ (black dashed histogram), and $\Delta\gamma^{\prime}>0.1$ (blue solid histogram), corresponding to the total density profile becoming `shallower', `invariant', and `steeper' as a result by AGN feedback energy in the given time intervals.

As shown in the figure, the distributions of $\Delta E_{\mathrm{Kin}}$, $\Delta E_{\mathrm{Thm}}$, and $\Delta\gamma^{\prime}$ are similar during $z\in[0, 0.5]$ and $z\in[0.5, 1]$, while being significantly different during $z\in[1, 2.5]$ from the two lower redshift bins. In the redshift range $z\in[1, 2.5]$ (bottom panel), the kinetic mode feedback energy span a larger range of values than the thermal mode feedback. From the color coding of the scattered dots, an excess of shallower $\Delta\gamma^{\prime}$ is present at both high $\Delta E_{\mathrm{Kin}}$ and high $\Delta E_{\mathrm{Thm}}$, indicating AGN feedback indeed makes the profile shallower. We point out that for galaxies which have relatively lower thermal feedback rates of $\Delta E_{\mathrm{Thm}}\sim 10^{53}\ \mathrm{ergs}$, the difference in $\Delta\gamma^{\prime}$ is mainly determined by their kinetic feedback rate $\Delta E_{\mathrm{Kin}}$. This reflects the higher feedback efficiency of the kinetic mode compared to the thermal mode in the quiescent AGN phase, and the low accretion rate kinetic AGN feedback couples more efficiently to the surrounding gas by driving shocks~\citep{2017MNRAS.465.3291W}. This is consistent with the finding in Section 4 of Paper I that removing the kinetic mode feedback and allowing the thermal mode to act at all black hole accretion rates could increase the overall distribution of $\gamma^{\prime}$ by $\sim 0.3$, which emphasizes the importance of invoking the kinetic mode AGN feedback to better match the observed galaxy correlations of $\gamma^{\prime}$. In addition, most of the main progenitors still become shallower in this period (red histogram), consistent with the overall redshift evolution trend of $\gamma^{\prime}$ becoming shallower during $z\in[1,2]$ as shown in Fig.~\ref{fig:tracer_TOT}. In the redshift ranges $[0.5, 1]$ and $[0, 0.5]$, the feedback energy deposited by the kinetic mode is comparable to that of the thermal mode, and there is no significant difference of the impact on $\Delta\gamma^{\prime}$ for the two modes. The bi-modal distribution of $\Delta\gamma^{\prime}$ seen at $z\in[1, 2.5]$ disappears, while the change in the total power-law density profile is dominated by invariance ($\Delta\gamma^{\prime}\in[-0.1, 0.1]$) in these two redshift ranges, which is shown by both the color coded scattered dots as well as the projected $\Delta\gamma^{\prime}-\Delta E_{\mathrm{Kin}}$ and $\Delta\gamma^{\prime}-\Delta E_{\mathrm{Thm}}$ distributions. This is also consistent with the redshift evolution trend of $\gamma^{\prime}$ for $z\leqslant 1$ as shown by Fig.~\ref{fig:tracer_TOT}. Appendix \ref{sec: B} shows further analyses of the statistical significance between AGN feedback and the change of the total density profile. 
 
The difference of the impact of the AGN feeback energy on the total density profile in different redshifts, i.e. $z > 1$ and $z\leqslant 1$, is mainly due to the sample selection we have applied. Our MPB sample is selected such that all galaxies have turned into ETGs by $z=1$, so the effect of AGN feedback energy on quenching star formation by removing baryons from the galactic central region (kinetic mode, e.g. \citealt{2013MNRAS.432.1947M}) and heating the surrounding interstellar medium (thermal mode) is more significant before $z=1$. The bottom panel ($z\in[1, 2.5]$) in Fig.~\ref{fig:bh_tracks} corresponds to the middle panel ($z\in[1,2]$) in Fig~\ref{fig:merger}, where the effects of major dry mergers compete with the rapid minor wet mergers, making AGN feedback energy (especially in the kinetic mode) more efficient at altering the total density profile. The depletion of gas in the main progenitors below $z=1$ (ETGs by then) diminishes the `working surface' for AGN feedback effects and makes the impact of AGN feedback energy on the evolution of $\gamma^{\prime}$ sub-dominant (upper two panels in Fig.~\ref{fig:bh_tracks}) compared to the impact of gas-poor mergers ($f_{\mathrm{Cold}}\lesssim 0.4$) in this redshift range (top panel in Fig~\ref{fig:merger}).
 
\subsection{A formation path for isothermal total density profiles}
\label{sec: 4.3}
 
We summarize the analysis of the MPB sample total density profile evolution as follows: the main progenitors start out with shallower than isothermal ($\gamma^{\prime}=2$) total density profiles at $z=4$ that have large scatter; \textbf{i)} during $z\in[2, 4]$, rapid wet mergers dominate the evolutionary trend of $\gamma^{\prime}$, making the total density profile steeper with time through dissipation processes; \textbf{ii)} $\gamma^{\prime}$ peaks at $z\approx2$ and decreases with time during $z\in[1, 2]$, with AGN feedback energy (especially through the kinetic mode) dominating the change of the total density profile, while rapid minor wet mergers compete with the effect of major dry mergers; \textbf{iii)} the main progenitors quench and turn into ETGs at $z=1$, followed by a passive evolution scenario during $z\in[0, 1]$ with almost invariant $\gamma^{\prime}$, which is dominated by gas-poor mergers, before reaching a near-isothermal (slightly shallower) total density profile at $z=0$.
 
With this analysis, we have identified a formation path of isothermality in ETGs that closely relates the evolution of the total density profile to the major activities that are crucial to ETG formation. In addition to the in-situ star formation and wet merger dominated (phase \textbf{i}), and dry merger dominated (phase \textbf{iii}) phases of the conventional two-phase ETG formation picture (see references in Section~\ref{sec: 1}), we emphasize the role of AGN feedback processes (especially in the kinetic mode) in $z\approx2.5-1$ (phase \textbf{ii}), making the total density profile shallower at around the transition (quenching) time of the selected ETG sample. Together, the three phases evolve ETG progenitors that span an order-of-magnitude in stellar mass and largely scatter their initial $\gamma^{\prime}$ towards a final isothermal state for their matter distribution. This highlights the predictive power of the IllustrisTNG Simulations for proposing a plausible evolutionary path that can explain the origin of the ETG `bulge-halo conspiracy'. 

Nonetheless, this scenario is closely related to the AGN feedback model adopted in the IllustrisTNG Simulations, which is known to possess some limitations that expect future improvements. Specifically, the total density profile of the MPB sample at $z=0$ in this work is shallower than observations (see Table A1 in\ Paper I). With the discrepancies in halo contraction and `density slope-velocity dispersion' relation compared to observations presented in Paper I, these findings might point to common limitations in the kinetic mode feedback in the IllustrisTNG AGN model. A census of the cool-cores in IllustrisTNG galaxy clusters also suggested an overly violent thermal mode feedback in the adopted IllustrisTNG AGN model~\citep{2018MNRAS.481.1809B}. Alternative black hole growth models, e.g. galaxy-scale torque-limited model~\citep{2013ApJ...770....5A,2015ApJ...800..127A,2017MNRAS.464.2840A,2017MNRAS.472L.109A}, may weaken the impact of AGN feedback on galaxy scale properties including the total density profile. This is not to undermine the great success of the kinetic wind AGN feedback in the IllustrisTNG model that mitigate the major tensions of the original Illustris results with observations, rather, it provides a valuable pivot point for further refinement of the current AGN model. Thus, future improvements in the black hole growth and AGN feedback models such as including the evolution of black hole spin~\citep{2019MNRAS.490.4133} are crucial for further constraining this multi-phase formation path of ETG isothermal density profiles.

\section{Conclusions and Discussion}
\label{sec: 5}

In this work we have focused on the evolutionary trend of the total density profiles in early-type galaxies selected from the state-of-the-art cosmological hydrodynamic simulation IllustrisTNG~\citep{2018MNRAS.480.5113M,2018MNRAS.477.1206N,2018MNRAS.475..624N,2018MNRAS.475..648P,2018MNRAS.475..676S}. Our analysis is broadly based on the realistic sample of IllustrisTNG ETGs studied in Paper I of the series~\citep{2018arXiv181106545W}, which provides a good match to the various correlations between the total power-law density slope $\gamma^{\prime}$ and galaxy properties (i.e. stellar mass, effective radius, central dark matter fraction etc.) seen in observations. Galaxy morphology classification is achieved utilizing single and double component fits to the SDSS $r-$band luminosity profile in each simulated galaxy that has best-fit S$\mathrm{\acute{e}}$rsic index $n\geqslant2$. With this classification criterion, we select ETGs at $z=0$ with stellar mass enclosed within their central $30\,\mathrm{kpc}$ that satisfy $\mathrm{log}\,(M_{\ast}/\mathrm{M_{\astrosun}})\in[10.7, 11.9]$, and that already have transformed to an ETG before $z=1$ by tracing their evolutionary tracks along the main progenitor branch of the baryonic version of the \textsc{sublink}~\citep{2015MNRAS.449...49R} merger tree. This results in our \textbf{main progenitor branch} sample that consists of 165 ETGs at $z=0$ with their individual evolution history traced out to $z=4$. We also select a \textbf{statistical ETG sample} that resembles observational ETG targets by applying the same stellar mass cut and ETG criteria in all selected redshifts from $z=0$ to $z=3$ for comparison (almost identical to the ETG samples shown in Fig. 11 of Paper I). The measurement of the galaxy total density profile is done using the best-fit total power-law density slope $\gamma^{\prime}$ within $[0.4\,R_{1/2}, 4\,R_{1/2}]$, where $R_{1/2}$ is the 3D stellar half mass radius of each galaxy. With an analysis of the total density profile depending on stellar mass and quenching redshift, as well as the impact of galaxy mergers and AGN feedback, we have identified a \textbf{multi-phase formation path} for the near-isothermal total density profiles in ETGs. We summarize our main findings as follows:

$\bullet$ The stellar mass evolution of the main progenitors roughly resembles that of a typical two-phase formation scenario, i.e. a quick rise from $z=4$ to $z\approx 1$ due to in-situ star formation/gas-rich mergers, and passive evolution via dry mergers with little increase in stellar mass from $z\approx 1$ to $z=0$. The sample is divided into 4 stellar mass bins based on their $z=0$ stellar mass in the central $30\,\mathrm{kpc}$, i.e. $\mathrm{log}\,(M_{\ast}/\mathrm{M_{\astrosun}})\in [10.7, 10.9], [10.9, 11.1], [11.1, 11.4], [11.4, 11.9]$ (Fig.~\ref{fig:m_ev}). The statistical ETG sample is selected to cover the same four stellar mass bins in all selected redshifts, which are dominated by lower stellar mass ETGs and have increasing proportions of them towards lower redshift (Fig.~\ref{fig:stats_mev}).

$\bullet$ The power-law slope of the total density profile of the MPB sample rises from a shallower-than-isothermal state at $z=4$, to a steeper-than-isothermal state peak value at $z\approx 2$, decreases steadfastly from $z=2$ to $z=1$, and finally decreases slightly with almost no evolution below $z=1$. The statistical ETG sample's $\gamma^{\prime}$ decreases quickly from $z=3$ to $z=1$, and is almost constant below $z=1$, although a mild increase in $\gamma^{\prime}$ is found from $z=0.5$ to $z=0$ (Fig.~\ref{fig:tracer_TOT}). We note that the $\gamma^{\prime}$ evolution of both samples have large scatter, and their differences are subject to statistical uncertainties. 

$\bullet$ The stellar mass dependence of the total density profile evolution is more significant for the main progenitors compared to the statistical ETG sample. The progenitors of the more (less) massive galaxies at $z=0$ in the MPB sample set out with steeper (shallower) seed $\gamma^{\prime}$ at $z=4$, rises quicker (slower) to the peak value of $\gamma^{\prime}$, and becomes shallower with time until $z=0$. The lower mass samples' $\gamma^{\prime}$ are closer to isothermal and steeper than the total density profile of their counterparts, consistent with the negative $\gamma^{\prime}-\mathrm{log}\,M_{\ast}$ correlation found in Paper I. The stellar mass dependence of the statistical ETG sample $\gamma^{\prime}$ evolution is less significant, mainly decreasing with time above $z=1$, and remaining invariant under $z=1$. The slight increase in $\gamma^{\prime}$ from $z=0.5$ to $z=0$ is more significant in lower mass bins. The scatter in the $\gamma^{\prime}$ evolution decreases with increasing stellar mass in the four bins, and the deviation from an isothermal total density profile is significant for the two higher stellar mass bins towards low redshift (Fig.~\ref{fig:tracer_bins}).

$\bullet$ Making variations to the MPB sample by changing the quenching redshift where all progenitors have turned to ETGs, we analyze the effect of different quenching redshifts on the $z=0$ $\gamma^{\prime}$ distribution. The sample's $z=0$ $\gamma^{\prime}$ median increases with decreasing quenching redshift. The corresponding descendant ETG sample at $z=0$ also involves more lower stellar mass galaxies with lower quenching redshift. This indicates that non-negligible progenitor bias could partially account for the apparent increase in $\gamma^{\prime}$ below $z=0.5$ for the statistical ETG sample (Fig.~\ref{fig:quench}). 

$\bullet$ The impact of galaxy mergers on $\gamma^{\prime}$ is shown in Fig.~\ref{fig:merger}. During $z\in[2, 4]$, mergers are dominated by minor wet mergers. These mergers mostly induce steeper density profiles as a consequence. During $z\in[1, 2]$, major mergers are mainly dry, while minor mergers still tend to be gas-rich. The major dry mergers are related to making the total density profile shallower, and minor gas-rich mergers continue to make the density profile steeper. Their effects counteract each other in this redshift range. During $z\in[0, 1]$, the mergers are dominated by gas-poor events, that induce shallower $\gamma^{\prime}$ as a consequence. The role of merger mass ratio is sub-dominant compared with the cold gas fraction in the merger, which marks the amount of dissipation that alters the total density profile. In fact, the correlation between the merger mass ratio and cold gas fraction (although weak) is a consequence of more massive infall progenitors in the major mergers that evolved faster and quenched earlier than their less massive counterparts in minor mergers. With major mergers being dry inducing shallower total density profile, the positive  $\gamma^{\prime}-f_{\mathrm{in-situ}}$ correlation at $z=0$ for IllustrisTNG ETGs is a result of the dominant role that dry major mergers play at lower redshift in decreasing $f_{\mathrm{in-situ}}$ and $\gamma^{\prime}$ simultaneously. 

$\bullet$ The impact of AGN feedback on $\gamma^{\prime}$ is shown in Fig.~\ref{fig:bh_tracks}. During $z\in[1, 2.5]$, the kinetic mode feedback energy displays a more significant bi-modal impact on the change of $\gamma^{\prime}$ compared to the thermal mode, with larger (smaller) kinetic mode feedback energy corresponding to shallower (steeper) total density profile. This suggests a dominant role of kinetic feedback removing baryons from galaxy central regions that effectively evolves the total density profile shallower with time. The effect of AGN feedback in both the kinetic mode and the thermal mode are consistent with no correlation to $\gamma^{\prime}$ during $z\in[0, 1]$, with both the thermal and kinetic modes inducing nearly constant $\gamma^{\prime}$ in this period.

$\bullet$ The \textbf{multi-phase formation path} of the isothermal total density profile in ETGs consists of: first, rapid wet mergers tend to steepen the total density profile through dissipation process from $z=4$ to $z=2$; second, effects of dry-major and wet-minor mergers from $z=2$ to $z=1$ roughly balance out, AGN feedback (especially kinetic mode) dominates the change of $\gamma^{\prime}$ by removing baryons from the galaxy central region; third, from $z=1$ to $z=0$, main progenitors passively evolve mainly through gas-poor mergers that mildly decrease $\gamma^{\prime}$ while maintaining the galaxy mass distribution close to isothermal. The main addition of this scenario to the conventional two-phase ETG formation picture is the role of AGN feedback making the total density profile shallower at around the type-transition time of galaxy evolution.

Our analysis is the first attempt to study a large sample of ETGs and trace their evolutionary tracks along the main progenitor branch of their merger trees for the evolution of their total density profiles in a state-of-the-art cosmological hydrodynamical simulation. We emphasize that our multi-phase formation path of isothermal-$\gamma^{\prime}$ is the first numerical evidence of the conventional `two-phase' ETG formation scenario being directly linked to the emergence of the `bulge-halo conspiracy', and that the formation path of isothermality in ETGs is also mass-dependent. This work extends the redshift evolution of $\gamma^{\prime}$ to higher redshift and compensates the statistical-sample-only redshift evolution studies on $\gamma^{\prime}$ in Illustris~\citep{2017MNRAS.469.1824X}, the Magneticum Pathfinder~\citep{2017MNRAS.464.3742R}, and Horizon-AGN~\citep{2017MNRAS.472.2153P,2019MNRAS.483.4615P}.  We find a mildly decreasing trend of $\gamma^{\prime}$ below $z=1$ for the main progenitors and a nearly-constant trend for the statistical sample, consistent with the findings of \citet{2017MNRAS.469.1824X} and \citet{2017MNRAS.464.3742R}. However, \citet{2017MNRAS.472.2153P} suggests that the weakening of AGN feedback at lower redshifts leads to steeper dark matter profiles. Combined with the nearly-invariant stellar profiles, this leads to a continuously steepening $\gamma^{\prime}$ below $z=2$ which better matches the observed redshift evolution trend of $\gamma^{\prime}$~\citep{2011ApJ...727...96R,2012ApJ...757...82B,2013ApJ...777...98S,2014MNRAS.440.2013D,2017ApJ...840...34S,2018MNRAS.480..431L}. Their slope values are smaller and their ETG sizes are somewhat more extended compared with observations due to AGN feedback. The major difference that we find for the IllustrisTNG ETGs is that they evolve shallower from a $\gamma^{\prime}\sim 2.2$ at $z\approx 2$ through a combination of AGN feedback and gas-poor mergers to the near-isothermal values at $z=0$, whereas the Horizon-AGN ETGs start out from $\gamma^{\prime}\sim 1.6$ at $z=2$ and continuously steepens until the near-isothermal values at $z=0$. These differences suggest that different AGN feedback and subgrid prescriptions result in different evolution paths for $\gamma^{\prime}$, and the coincidentally convergent near-isothermal values of $\gamma^{\prime}$ at $z=0$ call for future observations to better constrain the total density profiles of $z\gtrsim2$ ETGs to discern the real evolution of $\gamma^{\prime}$ in ETGs.

Apart from the discrepancies in halo contraction and $\gamma^{\prime}-\sigma_{\mathrm{e/2}}$ compared to observations found in Paper I, the total density profile of the MPB sample at $z=0$ in this work is slightly shallower compared with observations (see Table A1 in\ Paper I). These add up to suggest an underlying overly efficient implementation of the kinetic mode AGN feedback in the IllustrisTNG model. A study of the cool-cores in IllustrisTNG galaxy clusters also suggested an overly violent thermal mode feedback in the adopted IllustrisTNG AGN model~\citep{2018MNRAS.481.1809B}. Furthermore, \citet{2019arXiv190101095M} recently found a counter-intuitive phenomenon that stronger stellar and AGN feedback leads to unrealistically steeper $\gamma^{\prime}$ for mock strong lensing galaxies in the EAGLE Simulations~\citep{2015MNRAS.446..521S,2015MNRAS.450.1937C}. Galaxy-scale torque-limited black hole growth model~\citep{2013ApJ...770....5A,2015ApJ...800..127A,2017MNRAS.464.2840A,2017MNRAS.472L.109A} or alternative BH growth models may even weaken the impact of AGN feedback on galaxy scale properties such as the total density profile. Thus, more detailed observations and theoretical models for AGN are required to work towards a fully consistent picture of the role of AGN feedback in galaxy formation.

Apart from these limitations, \citet{2018MNRAS.475.2878S} reported fast and sharp increases of the S$\mathrm{\acute{e}}$rsic index with increasing redshift in observations, and it is unclear whether this is feasible in the hierarchical assembly context of galaxy formation. Galaxy mergers which are concurrent with AGN feedback processes also require more detailed modelling of merger mass ratios, cold gas fraction, and resolved infall progenitor properties to gain a more detailed view of the merger impact on the evolution of the total density profile, as well as ETG formation in general. Future improvements in the AGN feedback and other simulation models that could better reconcile the discrepancies aforementioned are crucial to better understanding the origin of the bulge-halo conspiracy.

\section*{Acknowledgements}

We thank Annalisa Pillepich, Daniel Eisenstein, Dylan Nelson, Rhea-Silvia Remus, Ryan McKinnon, and Stephanie O'Neil for helpful discussions and support during the preparation of this paper. We thank the anonymous referee for carefully reading the draft and providing insightful comments that helped to improve this paper. YW acknowledges the Tsinghua Xuetang Talents Programme for funding his research at MIT. MV acknowledges support through an MIT RSC award, a Kavli Research Investment Fund, NASA ATP grant NNX17AG29G, and NSF grants AST-1814053 and AST-1814259. DX would like to thank the supercomputing facilities at the Heidelberg Institute for Advanced Studies and the Klaus Tschira Foundation. This work is partly supported by a joint grant between the DFG and NSFC (Grant No. 11761131004), the National Key Basic Research and Development Programme of China (No. 2018YFA0404501), and grant 11761131004 of NSFC. FM acknowledges support through the Programme `Rita Levi Montalcini' of the Italian MIUR.




\bibliographystyle{mnras}
\bibliography{isothermal2} 

\begin{thebibliography}{}
\makeatletter
\relax
\def\mn@urlcharsother{\let\do\@makeother \do\$\do\&\do\#\do\^\do\_\do\%\do\~}
\def\mn@doi{\begingroup\mn@urlcharsother \@ifnextchar [ {\mn@doi@}
  {\mn@doi@[]}}
\def\mn@doi@[#1]#2{\def\@tempa{#1}\ifx\@tempa\@empty \href
  {http://dx.doi.org/#2} {doi:#2}\else \href {http://dx.doi.org/#2} {#1}\fi
  \endgroup}
\def\mn@eprint#1#2{\mn@eprint@#1:#2::\@nil}
\def\mn@eprint@arXiv#1{\href {http://arxiv.org/abs/#1} {{\tt arXiv:#1}}}
\def\mn@eprint@dblp#1{\href {http://dblp.uni-trier.de/rec/bibtex/#1.xml}
  {dblp:#1}}
\def\mn@eprint@#1:#2:#3:#4\@nil{\def\@tempa {#1}\def\@tempb {#2}\def\@tempc
  {#3}\ifx \@tempc \@empty \let \@tempc \@tempb \let \@tempb \@tempa \fi \ifx
  \@tempb \@empty \def\@tempb {arXiv}\fi \@ifundefined
  {mn@eprint@\@tempb}{\@tempb:\@tempc}{\expandafter \expandafter \csname
  mn@eprint@\@tempb\endcsname \expandafter{\@tempc}}}

\bibitem[\protect\citeauthoryear{{Angl{\'e}s-Alc{\'a}zar}, {{\"O}zel}  \&
  {Dav{\'e}}}{{Angl{\'e}s-Alc{\'a}zar} et~al.}{2013}]{2013ApJ...770....5A}
{Angl{\'e}s-Alc{\'a}zar} D.,  {{\"O}zel} F.,   {Dav{\'e}} R.,  2013, \mn@doi
  [\apj] {10.1088/0004-637X/770/1/5}, \href
  {http://adsabs.harvard.edu/abs/2013ApJ...770....5A} {770, 5}

\bibitem[\protect\citeauthoryear{{Angl{\'e}s-Alc{\'a}zar}, {{\"O}zel},
  {Dav{\'e}}, {Katz}, {Kollmeier}  \& {Oppenheimer}}{{Angl{\'e}s-Alc{\'a}zar}
  et~al.}{2015}]{2015ApJ...800..127A}
{Angl{\'e}s-Alc{\'a}zar} D.,  {{\"O}zel} F.,  {Dav{\'e}} R.,  {Katz} N.,
  {Kollmeier} J.~A.,   {Oppenheimer} B.~D.,  2015, \mn@doi [\apj]
  {10.1088/0004-637X/800/2/127}, \href
  {http://adsabs.harvard.edu/abs/2015ApJ...800..127A} {800, 127}

\bibitem[\protect\citeauthoryear{{Angl{\'e}s-Alc{\'a}zar}, {Dav{\'e}},
  {Faucher-Gigu{\`e}re}, {{\"O}zel}  \& {Hopkins}}{{Angl{\'e}s-Alc{\'a}zar}
  et~al.}{2017a}]{2017MNRAS.464.2840A}
{Angl{\'e}s-Alc{\'a}zar} D.,  {Dav{\'e}} R.,  {Faucher-Gigu{\`e}re} C.-A.,
  {{\"O}zel} F.,   {Hopkins} P.~F.,  2017a, \mn@doi [\mnras]
  {10.1093/mnras/stw2565}, \href
  {http://adsabs.harvard.edu/abs/2017MNRAS.464.2840A} {464, 2840}

\bibitem[\protect\citeauthoryear{{Angl{\'e}s-Alc{\'a}zar},
  {Faucher-Gigu{\`e}re}, {Quataert}, {Hopkins}, {Feldmann}, {Torrey}, {Wetzel}
  \& {Kere{\v s}}}{{Angl{\'e}s-Alc{\'a}zar}
  et~al.}{2017b}]{2017MNRAS.472L.109A}
{Angl{\'e}s-Alc{\'a}zar} D.,  {Faucher-Gigu{\`e}re} C.-A.,  {Quataert} E.,
  {Hopkins} P.~F.,  {Feldmann} R.,  {Torrey} P.,  {Wetzel} A.,   {Kere{\v s}}
  D.,  2017b, \mn@doi [\mnras] {10.1093/mnrasl/slx161}, \href
  {http://adsabs.harvard.edu/abs/2017MNRAS.472L.109A} {472, L109}

\bibitem[\protect\citeauthoryear{{Auger}, {Treu}, {Bolton}, {Gavazzi},
  {Koopmans}, {Marshall}, {Moustakas}  \& {Burles}}{{Auger}
  et~al.}{2010}]{2010ApJ...724..511A}
{Auger} M.~W.,  {Treu} T.,  {Bolton} A.~S.,  {Gavazzi} R.,  {Koopmans}
  L.~V.~E.,  {Marshall} P.~J.,  {Moustakas} L.~A.,   {Burles} S.,  2010,
  \mn@doi [\apj] {10.1088/0004-637X/724/1/511}, \href
  {http://adsabs.harvard.edu/abs/2010ApJ...724..511A} {724, 511}

\bibitem[\protect\citeauthoryear{{Barnab{\`e}}, {Czoske}, {Koopmans}, {Treu},
  {Bolton}  \& {Gavazzi}}{{Barnab{\`e}} et~al.}{2009}]{2009MNRAS.399...21B}
{Barnab{\`e}} M.,  {Czoske} O.,  {Koopmans} L.~V.~E.,  {Treu} T.,  {Bolton}
  A.~S.,   {Gavazzi} R.,  2009, \mn@doi [\mnras]
  {10.1111/j.1365-2966.2009.14941.x}, \href
  {http://adsabs.harvard.edu/abs/2009MNRAS.399...21B} {399, 21}

\bibitem[\protect\citeauthoryear{{Barnab{\`e}}, {Czoske}, {Koopmans}, {Treu}
  \& {Bolton}}{{Barnab{\`e}} et~al.}{2011}]{2011MNRAS.415.2215B}
{Barnab{\`e}} M.,  {Czoske} O.,  {Koopmans} L.~V.~E.,  {Treu} T.,   {Bolton}
  A.~S.,  2011, \mn@doi [\mnras] {10.1111/j.1365-2966.2011.18842.x}, \href
  {http://adsabs.harvard.edu/abs/2011MNRAS.415.2215B} {415, 2215}

\bibitem[\protect\citeauthoryear{{Barnes} et~al.,}{{Barnes}
  et~al.}{2018}]{2018MNRAS.481.1809B}
{Barnes} D.~J.,  et~al., 2018, \mn@doi [\mnras] {10.1093/mnras/sty2078}, \href
  {http://adsabs.harvard.edu/abs/2018MNRAS.481.1809B} {481, 1809}

\bibitem[\protect\citeauthoryear{{Bellstedt} et~al.,}{{Bellstedt}
  et~al.}{2018}]{2018MNRAS.476.4543B}
{Bellstedt} S.,  et~al., 2018, \mn@doi [\mnras] {10.1093/mnras/sty456}, \href
  {http://adsabs.harvard.edu/abs/2018MNRAS.476.4543B} {476, 4543}

\bibitem[\protect\citeauthoryear{{Bluck} et~al.,}{{Bluck}
  et~al.}{2016}]{2016MNRAS.462.2559B}
{Bluck} A.~F.~L.,  et~al., 2016, \mn@doi [\mnras] {10.1093/mnras/stw1665},
  \href {http://adsabs.harvard.edu/abs/2016MNRAS.462.2559B} {462, 2559}

\bibitem[\protect\citeauthoryear{{Blumenthal}, {Faber}, {Primack}  \&
  {Rees}}{{Blumenthal} et~al.}{1984}]{1984Natur.311..517B}
{Blumenthal} G.~R.,  {Faber} S.~M.,  {Primack} J.~R.,   {Rees} M.~J.,  1984,
  \mn@doi [\nat] {10.1038/311517a0}, \href
  {http://adsabs.harvard.edu/abs/1984Natur.311..517B} {311, 517}

\bibitem[\protect\citeauthoryear{{Bolton} et~al.,}{{Bolton}
  et~al.}{2012}]{2012ApJ...757...82B}
{Bolton} A.~S.,  et~al., 2012, \mn@doi [\apj] {10.1088/0004-637X/757/1/82},
  \href {http://adsabs.harvard.edu/abs/2012ApJ...757...82B} {757, 82}

\bibitem[\protect\citeauthoryear{{Booth} \& {Schaye}}{{Booth} \&
  {Schaye}}{2009}]{2009MNRAS.398...53B}
{Booth} C.~M.,  {Schaye} J.,  2009, \mn@doi [\mnras]
  {10.1111/j.1365-2966.2009.15043.x}, \href
  {https://ui.adsabs.harvard.edu/abs/2009MNRAS.398...53B} {398, 53}

\bibitem[\protect\citeauthoryear{{Bower}, {Schaye}, {Frenk}, {Theuns},
  {Schaller}, {Crain}  \& {McAlpine}}{{Bower}
  et~al.}{2017}]{2017MNRAS.465...32B}
{Bower} R.~G.,  {Schaye} J.,  {Frenk} C.~S.,  {Theuns} T.,  {Schaller} M.,
  {Crain} R.~A.,   {McAlpine} S.,  2017, \mn@doi [\mnras]
  {10.1093/mnras/stw2735}, \href
  {http://adsabs.harvard.edu/abs/2017MNRAS.465...32B} {465, 32}

\bibitem[\protect\citeauthoryear{{Brownstein} et~al.,}{{Brownstein}
  et~al.}{2012}]{2012ApJ...744...41B}
{Brownstein} J.~R.,  et~al., 2012, \mn@doi [\apj] {10.1088/0004-637X/744/1/41},
  \href {http://adsabs.harvard.edu/abs/2012ApJ...744...41B} {744, 41}

\bibitem[\protect\citeauthoryear{{Bruzual} \& {Charlot}}{{Bruzual} \&
  {Charlot}}{2003}]{2003MNRAS.344.1000B}
{Bruzual} G.,  {Charlot} S.,  2003, \mn@doi [\mnras]
  {10.1046/j.1365-8711.2003.06897.x}, \href
  {http://adsabs.harvard.edu/abs/2003MNRAS.344.1000B} {344, 1000}

\bibitem[\protect\citeauthoryear{{Bundy} et~al.,}{{Bundy}
  et~al.}{2015}]{2015ApJ...798....7B}
{Bundy} K.,  et~al., 2015, \mn@doi [\apj] {10.1088/0004-637X/798/1/7}, \href
  {https://ui.adsabs.harvard.edu/abs/2015ApJ...798....7B} {798, 7}

\bibitem[\protect\citeauthoryear{{Bustamante} \& {Springel}}{{Bustamante} \&
  {Springel}}{2019}]{2019MNRAS.490.4133}
{Bustamante} S.,  {Springel} V.,  2019, \mn@doi [\mnras]
  {10.1093/mnras/stz2836}, \href
  {https://ui.adsabs.harvard.edu/abs/2019MNRAS.tmp.2455B} {490, 4133}

\bibitem[\protect\citeauthoryear{{Cappellari}}{{Cappellari}}{2016}]{2016ARA&A..54..597C}
{Cappellari} M.,  2016, \mn@doi [\araa] {10.1146/annurev-astro-082214-122432},
  \href {http://adsabs.harvard.edu/abs/2016ARA%26A..54..597C} {54, 597}

\bibitem[\protect\citeauthoryear{{Cappellari} et~al.,}{{Cappellari}
  et~al.}{2011}]{2011MNRAS.413..813C}
{Cappellari} M.,  et~al., 2011, \mn@doi [\mnras]
  {10.1111/j.1365-2966.2010.18174.x}, \href
  {http://adsabs.harvard.edu/abs/2011MNRAS.413..813C} {413, 813}

\bibitem[\protect\citeauthoryear{{Cappellari} et~al.,}{{Cappellari}
  et~al.}{2013}]{2013MNRAS.432.1709C}
{Cappellari} M.,  et~al., 2013, \mn@doi [\mnras] {10.1093/mnras/stt562}, \href
  {http://adsabs.harvard.edu/abs/2013MNRAS.432.1709C} {432, 1709}

\bibitem[\protect\citeauthoryear{{Cappellari} et~al.,}{{Cappellari}
  et~al.}{2015}]{2015ApJ...804L..21C}
{Cappellari} M.,  et~al., 2015, \mn@doi [\apjl] {10.1088/2041-8205/804/1/L21},
  \href {http://adsabs.harvard.edu/abs/2015ApJ...804L..21C} {804, L21}

\bibitem[\protect\citeauthoryear{{Cheung} et~al.,}{{Cheung}
  et~al.}{2016}]{2016Natur.533..504C}
{Cheung} E.,  et~al., 2016, \mn@doi [Nature] {10.1038/nature18006}, \href
  {https://ui.adsabs.harvard.edu/abs/2016Natur.533..504C} {533, 504}

\bibitem[\protect\citeauthoryear{{Ciotti} \& {van Albada}}{{Ciotti} \& {van
  Albada}}{2001}]{2001ApJ...552L..13C}
{Ciotti} L.,  {van Albada} T.~S.,  2001, \mn@doi [\apjl] {10.1086/320260},
  \href {http://adsabs.harvard.edu/abs/2001ApJ...552L..13C} {552, L13}

\bibitem[\protect\citeauthoryear{{Ciotti}, {Morganti}  \& {de Zeeuw}}{{Ciotti}
  et~al.}{2009}]{2009MNRAS.393..491C}
{Ciotti} L.,  {Morganti} L.,   {de Zeeuw} P.~T.,  2009, \mn@doi [\mnras]
  {10.1111/j.1365-2966.2008.14009.x}, \href
  {http://adsabs.harvard.edu/abs/2009MNRAS.393..491C} {393, 491}

\bibitem[\protect\citeauthoryear{{Ciotti}, {Pellegrini}, {Negri}  \&
  {Ostriker}}{{Ciotti} et~al.}{2017}]{2017ApJ...835...15C}
{Ciotti} L.,  {Pellegrini} S.,  {Negri} A.,   {Ostriker} J.~P.,  2017, \mn@doi
  [\apj] {10.3847/1538-4357/835/1/15}, \href
  {http://adsabs.harvard.edu/abs/2017ApJ...835...15C} {835, 15}

\bibitem[\protect\citeauthoryear{{Cole}, {Aragon-Salamanca}, {Frenk}, {Navarro}
   \& {Zepf}}{{Cole} et~al.}{1994}]{1994MNRAS.271..781C}
{Cole} S.,  {Aragon-Salamanca} A.,  {Frenk} C.~S.,  {Navarro} J.~F.,   {Zepf}
  S.~E.,  1994, \mn@doi [\mnras] {10.1093/mnras/271.4.781}, \href
  {http://adsabs.harvard.edu/abs/1994MNRAS.271..781C} {271, 781}

\bibitem[\protect\citeauthoryear{{Crain} et~al.,}{{Crain}
  et~al.}{2015}]{2015MNRAS.450.1937C}
{Crain} R.~A.,  et~al., 2015, \mn@doi [\mnras] {10.1093/mnras/stv725}, \href
  {http://adsabs.harvard.edu/abs/2015MNRAS.450.1937C} {450, 1937}

\bibitem[\protect\citeauthoryear{{De Lucia} \& {Blaizot}}{{De Lucia} \&
  {Blaizot}}{2007}]{2007MNRAS.375....2D}
{De Lucia} G.,  {Blaizot} J.,  2007, \mn@doi [\mnras]
  {10.1111/j.1365-2966.2006.11287.x}, \href
  {http://adsabs.harvard.edu/abs/2007MNRAS.375....2D} {375, 2}

\bibitem[\protect\citeauthoryear{{Deason}, {Belokurov}, {Evans}  \&
  {McCarthy}}{{Deason} et~al.}{2012}]{2012ApJ...748....2D}
{Deason} A.~J.,  {Belokurov} V.,  {Evans} N.~W.,   {McCarthy} I.~G.,  2012,
  \mn@doi [\apj] {10.1088/0004-637X/748/1/2}, \href
  {http://adsabs.harvard.edu/abs/2012ApJ...748....2D} {748, 2}

\bibitem[\protect\citeauthoryear{{Dekel} et~al.,}{{Dekel}
  et~al.}{2009}]{2009Natur.457..451D}
{Dekel} A.,  et~al., 2009, \mn@doi [\nat] {10.1038/nature07648}, \href
  {http://adsabs.harvard.edu/abs/2009Natur.457..451D} {457, 451}

\bibitem[\protect\citeauthoryear{{Di Matteo}, {Springel}  \& {Hernquist}}{{Di
  Matteo} et~al.}{2005}]{2005Natur.433..604D}
{Di Matteo} T.,  {Springel} V.,   {Hernquist} L.,  2005, \mn@doi [\nat]
  {10.1038/nature03335}, \href
  {http://adsabs.harvard.edu/abs/2005Natur.433..604D} {433, 604}

\bibitem[\protect\citeauthoryear{{Dolag}, {Borgani}, {Murante}  \&
  {Springel}}{{Dolag} et~al.}{2009}]{2009MNRAS.399..497D}
{Dolag} K.,  {Borgani} S.,  {Murante} G.,   {Springel} V.,  2009, \mn@doi
  [\mnras] {10.1111/j.1365-2966.2009.15034.x}, \href
  {http://adsabs.harvard.edu/abs/2009MNRAS.399..497D} {399, 497}

\bibitem[\protect\citeauthoryear{{Dubois}, {Devriendt}, {Slyz}  \&
  {Teyssier}}{{Dubois} et~al.}{2012}]{2012MNRAS.420.2662D}
{Dubois} Y.,  {Devriendt} J.,  {Slyz} A.,   {Teyssier} R.,  2012, \mn@doi
  [\mnras] {10.1111/j.1365-2966.2011.20236.x}, \href
  {https://ui.adsabs.harvard.edu/abs/2012MNRAS.420.2662D} {420, 2662}

\bibitem[\protect\citeauthoryear{{Dubois}, {Gavazzi}, {Peirani}  \&
  {Silk}}{{Dubois} et~al.}{2013}]{2013MNRAS.433.3297D}
{Dubois} Y.,  {Gavazzi} R.,  {Peirani} S.,   {Silk} J.,  2013, \mn@doi [\mnras]
  {10.1093/mnras/stt997}, \href
  {http://adsabs.harvard.edu/abs/2013MNRAS.433.3297D} {433, 3297}

\bibitem[\protect\citeauthoryear{{Dubois} et~al.,}{{Dubois}
  et~al.}{2014}]{2014MNRAS.444.1453D}
{Dubois} Y.,  et~al., 2014, \mn@doi [\mnras] {10.1093/mnras/stu1227}, \href
  {http://adsabs.harvard.edu/abs/2014MNRAS.444.1453D} {444, 1453}

\bibitem[\protect\citeauthoryear{{Duffy}, {Schaye}, {Kay}, {Dalla Vecchia},
  {Battye}  \& {Booth}}{{Duffy} et~al.}{2010}]{2010MNRAS.405.2161D}
{Duffy} A.~R.,  {Schaye} J.,  {Kay} S.~T.,  {Dalla Vecchia} C.,  {Battye}
  R.~A.,   {Booth} C.~M.,  2010, \mn@doi [\mnras]
  {10.1111/j.1365-2966.2010.16613.x}, \href
  {http://adsabs.harvard.edu/abs/2010MNRAS.405.2161D} {405, 2161}

\bibitem[\protect\citeauthoryear{{Dye} et~al.,}{{Dye}
  et~al.}{2014}]{2014MNRAS.440.2013D}
{Dye} S.,  et~al., 2014, \mn@doi [\mnras] {10.1093/mnras/stu305}, \href
  {http://adsabs.harvard.edu/abs/2014MNRAS.440.2013D} {440, 2013}

\bibitem[\protect\citeauthoryear{{Emsellem} et~al.,}{{Emsellem}
  et~al.}{2011}]{2011MNRAS.414..888E}
{Emsellem} E.,  et~al., 2011, \mn@doi [\mnras]
  {10.1111/j.1365-2966.2011.18496.x}, \href
  {http://adsabs.harvard.edu/abs/2011MNRAS.414..888E} {414, 888}

\bibitem[\protect\citeauthoryear{{Fabian}}{{Fabian}}{2012}]{2012ARA&A..50..455F}
{Fabian} A.~C.,  2012, \mn@doi [\araa] {10.1146/annurev-astro-081811-125521},
  \href {http://adsabs.harvard.edu/abs/2012ARA%26A..50..455F} {50, 455}

\bibitem[\protect\citeauthoryear{{Faucher-Gigu{\`e}re} \&
  {Quataert}}{{Faucher-Gigu{\`e}re} \& {Quataert}}{2012}]{2012MNRAS.425..605F}
{Faucher-Gigu{\`e}re} C.-A.,  {Quataert} E.,  2012, \mn@doi [\mnras]
  {10.1111/j.1365-2966.2012.21512.x}, \href
  {http://adsabs.harvard.edu/abs/2012MNRAS.425..605F} {425, 605}

\bibitem[\protect\citeauthoryear{{Franx}, {van Dokkum}, {F{\"o}rster
  Schreiber}, {Wuyts}, {Labb{\'e}}  \& {Toft}}{{Franx}
  et~al.}{2008}]{2008ApJ...688..770F}
{Franx} M.,  {van Dokkum} P.~G.,  {F{\"o}rster Schreiber} N.~M.,  {Wuyts} S.,
  {Labb{\'e}} I.,   {Toft} S.,  2008, \mn@doi [\apj] {10.1086/592431}, \href
  {http://adsabs.harvard.edu/abs/2008ApJ...688..770F} {688, 770}

\bibitem[\protect\citeauthoryear{{Furlong} et~al.,}{{Furlong}
  et~al.}{2015}]{2015MNRAS.450.4486F}
{Furlong} M.,  et~al., 2015, \mn@doi [\mnras] {10.1093/mnras/stv852}, \href
  {http://adsabs.harvard.edu/abs/2015MNRAS.450.4486F} {450, 4486}

\bibitem[\protect\citeauthoryear{{Gaspari}, {Brighenti}  \& {Temi}}{{Gaspari}
  et~al.}{2012}]{2012MNRAS.424..190G}
{Gaspari} M.,  {Brighenti} F.,   {Temi} P.,  2012, \mn@doi [\mnras]
  {10.1111/j.1365-2966.2012.21183.x}, \href
  {http://adsabs.harvard.edu/abs/2012MNRAS.424..190G} {424, 190}

\bibitem[\protect\citeauthoryear{{Gavazzi}, {Treu}, {Rhodes}, {Koopmans},
  {Bolton}, {Burles}, {Massey}  \& {Moustakas}}{{Gavazzi}
  et~al.}{2007}]{2007ApJ...667..176G}
{Gavazzi} R.,  {Treu} T.,  {Rhodes} J.~D.,  {Koopmans} L.~V.~E.,  {Bolton}
  A.~S.,  {Burles} S.,  {Massey} R.~J.,   {Moustakas} L.~A.,  2007, \mn@doi
  [\apj] {10.1086/519237}, \href
  {http://adsabs.harvard.edu/abs/2007ApJ...667..176G} {667, 176}

\bibitem[\protect\citeauthoryear{{Genel} et~al.,}{{Genel}
  et~al.}{2014}]{2014MNRAS.445..175G}
{Genel} S.,  et~al., 2014, \mn@doi [\mnras] {10.1093/mnras/stu1654}, \href
  {http://adsabs.harvard.edu/abs/2014MNRAS.445..175G} {445, 175}

\bibitem[\protect\citeauthoryear{{Genel}, {Fall}, {Hernquist}, {Vogelsberger},
  {Snyder}, {Rodriguez-Gomez}, {Sijacki}  \& {Springel}}{{Genel}
  et~al.}{2015}]{2015ApJ...804L..40G}
{Genel} S.,  {Fall} S.~M.,  {Hernquist} L.,  {Vogelsberger} M.,  {Snyder}
  G.~F.,  {Rodriguez-Gomez} V.,  {Sijacki} D.,   {Springel} V.,  2015, \mn@doi
  [\apjl] {10.1088/2041-8205/804/2/L40}, \href
  {http://adsabs.harvard.edu/abs/2015ApJ...804L..40G} {804, L40}

\bibitem[\protect\citeauthoryear{{Genel} et~al.,}{{Genel}
  et~al.}{2018}]{2018MNRAS.474.3976G}
{Genel} S.,  et~al., 2018, \mn@doi [\mnras] {10.1093/mnras/stx3078}, \href
  {http://adsabs.harvard.edu/abs/2018MNRAS.474.3976G} {474, 3976}

\bibitem[\protect\citeauthoryear{{Graham}}{{Graham}}{2008}]{2008ApJ...680..143G}
{Graham} A.~W.,  2008, \mn@doi [\apj] {10.1086/587473}, \href
  {http://adsabs.harvard.edu/abs/2008ApJ...680..143G} {680, 143}

\bibitem[\protect\citeauthoryear{{Guo} \& {White}}{{Guo} \&
  {White}}{2008}]{2008MNRAS.384....2G}
{Guo} Q.,  {White} S.~D.~M.,  2008, \mn@doi [\mnras]
  {10.1111/j.1365-2966.2007.12619.x}, \href
  {http://adsabs.harvard.edu/abs/2008MNRAS.384....2G} {384, 2}

\bibitem[\protect\citeauthoryear{{Hilz}, {Naab}, {Ostriker}, {Thomas},
  {Burkert}  \& {Jesseit}}{{Hilz} et~al.}{2012}]{2012MNRAS.425.3119H}
{Hilz} M.,  {Naab} T.,  {Ostriker} J.~P.,  {Thomas} J.,  {Burkert} A.,
  {Jesseit} R.,  2012, \mn@doi [\mnras] {10.1111/j.1365-2966.2012.21541.x},
  \href {http://adsabs.harvard.edu/abs/2012MNRAS.425.3119H} {425, 3119}

\bibitem[\protect\citeauthoryear{{Hilz}, {Naab}  \& {Ostriker}}{{Hilz}
  et~al.}{2013}]{2013MNRAS.429.2924H}
{Hilz} M.,  {Naab} T.,   {Ostriker} J.~P.,  2013, \mn@doi [\mnras]
  {10.1093/mnras/sts501}, \href
  {http://adsabs.harvard.edu/abs/2013MNRAS.429.2924H} {429, 2924}

\bibitem[\protect\citeauthoryear{{Hirschmann} et~al.,}{{Hirschmann}
  et~al.}{2013}]{2013MNRAS.436.2929H}
{Hirschmann} M.,  et~al., 2013, \mn@doi [\mnras] {10.1093/mnras/stt1770}, \href
  {http://adsabs.harvard.edu/abs/2013MNRAS.436.2929H} {436, 2929}

\bibitem[\protect\citeauthoryear{{Hopkins}, {Cox}, {Younger}  \&
  {Hernquist}}{{Hopkins} et~al.}{2009}]{2009ApJ...691.1168H}
{Hopkins} P.~F.,  {Cox} T.~J.,  {Younger} J.~D.,   {Hernquist} L.,  2009,
  \mn@doi [\apj] {10.1088/0004-637X/691/2/1168}, \href
  {http://adsabs.harvard.edu/abs/2009ApJ...691.1168H} {691, 1168}

\bibitem[\protect\citeauthoryear{{Hopkins}, {Torrey}, {Faucher-Gigu{\`e}re},
  {Quataert}  \& {Murray}}{{Hopkins} et~al.}{2016}]{2016MNRAS.458..816H}
{Hopkins} P.~F.,  {Torrey} P.,  {Faucher-Gigu{\`e}re} C.-A.,  {Quataert} E.,
  {Murray} N.,  2016, \mn@doi [\mnras] {10.1093/mnras/stw289}, \href
  {http://adsabs.harvard.edu/abs/2016MNRAS.458..816H} {458, 816}

\bibitem[\protect\citeauthoryear{{Humphrey} \& {Buote}}{{Humphrey} \&
  {Buote}}{2010}]{2010MNRAS.403.2143H}
{Humphrey} P.~J.,  {Buote} D.~A.,  2010, \mn@doi [\mnras]
  {10.1111/j.1365-2966.2010.16257.x}, \href
  {http://adsabs.harvard.edu/abs/2010MNRAS.403.2143H} {403, 2143}

\bibitem[\protect\citeauthoryear{{Humphrey}, {Buote}, {Gastaldello},
  {Zappacosta}, {Bullock}, {Brighenti}  \& {Mathews}}{{Humphrey}
  et~al.}{2006}]{2006ApJ...646..899H}
{Humphrey} P.~J.,  {Buote} D.~A.,  {Gastaldello} F.,  {Zappacosta} L.,
  {Bullock} J.~S.,  {Brighenti} F.,   {Mathews} W.~G.,  2006, \mn@doi [\apj]
  {10.1086/505019}, \href {http://adsabs.harvard.edu/abs/2006ApJ...646..899H}
  {646, 899}

\bibitem[\protect\citeauthoryear{{Johansson}, {Naab}  \& {Burkert}}{{Johansson}
  et~al.}{2009}]{2009ApJ...690..802J}
{Johansson} P.~H.,  {Naab} T.,   {Burkert} A.,  2009, \mn@doi [\apj]
  {10.1088/0004-637X/690/1/802}, \href
  {http://adsabs.harvard.edu/abs/2009ApJ...690..802J} {690, 802}

\bibitem[\protect\citeauthoryear{{Johansson}, {Naab}  \&
  {Ostriker}}{{Johansson} et~al.}{2012}]{2012ApJ...754..115J}
{Johansson} P.~H.,  {Naab} T.,   {Ostriker} J.~P.,  2012, \mn@doi [\apj]
  {10.1088/0004-637X/754/2/115}, \href
  {http://adsabs.harvard.edu/abs/2012ApJ...754..115J} {754, 115}

\bibitem[\protect\citeauthoryear{{Kere{\v s}}, {Katz}, {Weinberg}  \&
  {Dav{\'e}}}{{Kere{\v s}} et~al.}{2005}]{2005MNRAS.363....2K}
{Kere{\v s}} D.,  {Katz} N.,  {Weinberg} D.~H.,   {Dav{\'e}} R.,  2005, \mn@doi
  [\mnras] {10.1111/j.1365-2966.2005.09451.x}, \href
  {http://adsabs.harvard.edu/abs/2005MNRAS.363....2K} {363, 2}

\bibitem[\protect\citeauthoryear{{Khochfar} \& {Silk}}{{Khochfar} \&
  {Silk}}{2009}]{2009MNRAS.397..506K}
{Khochfar} S.,  {Silk} J.,  2009, \mn@doi [\mnras]
  {10.1111/j.1365-2966.2009.14958.x}, \href
  {http://adsabs.harvard.edu/abs/2009MNRAS.397..506K} {397, 506}

\bibitem[\protect\citeauthoryear{{King}}{{King}}{2003}]{2003ApJ...596L..27K}
{King} A.,  2003, \mn@doi [\apjl] {10.1086/379143}, \href
  {http://adsabs.harvard.edu/abs/2003ApJ...596L..27K} {596, L27}

\bibitem[\protect\citeauthoryear{{Koopmans}, {Treu}, {Bolton}, {Burles}  \&
  {Moustakas}}{{Koopmans} et~al.}{2006}]{2006ApJ...649..599K}
{Koopmans} L.~V.~E.,  {Treu} T.,  {Bolton} A.~S.,  {Burles} S.,   {Moustakas}
  L.~A.,  2006, \mn@doi [\apj] {10.1086/505696}, \href
  {http://adsabs.harvard.edu/abs/2006ApJ...649..599K} {649, 599}

\bibitem[\protect\citeauthoryear{{Koopmans} et~al.,}{{Koopmans}
  et~al.}{2009}]{2009ApJ...703L..51K}
{Koopmans} L.~V.~E.,  et~al., 2009, \mn@doi [\apjl]
  {10.1088/0004-637X/703/1/L51}, \href
  {http://adsabs.harvard.edu/abs/2009ApJ...703L..51K} {703, L51}

\bibitem[\protect\citeauthoryear{{Kormendy} \& {Ho}}{{Kormendy} \&
  {Ho}}{2013}]{2013ARA&A..51..511K}
{Kormendy} J.,  {Ho} L.~C.,  2013, \mn@doi [\araa]
  {10.1146/annurev-astro-082708-101811}, \href
  {http://adsabs.harvard.edu/abs/2013ARA%26A..51..511K} {51, 511}

\bibitem[\protect\citeauthoryear{{Krajnovi{\'c}} et~al.,}{{Krajnovi{\'c}}
  et~al.}{2011}]{2011MNRAS.414.2923K}
{Krajnovi{\'c}} D.,  et~al., 2011, \mn@doi [\mnras]
  {10.1111/j.1365-2966.2011.18560.x}, \href
  {http://adsabs.harvard.edu/abs/2011MNRAS.414.2923K} {414, 2923}

\bibitem[\protect\citeauthoryear{{Li H.} et~al.,}{{Li H.}
  et~al.}{2017}]{2017ApJ...838...77L}
{Li H.} Y.,  et~al., 2017, \mn@doi [\apj] {10.3847/1538-4357/aa662a}, \href
  {http://adsabs.harvard.edu/abs/2017ApJ...838...77L} {838, 77}

\bibitem[\protect\citeauthoryear{{Li H.}, {Mao}, {Emsellem}, {Xu}, {Springel}
  \& {Krajnovi{\'c}}}{{Li H.} et~al.}{2018a}]{2018MNRAS.473.1489L}
{Li H.} Y.,  {Mao} S.,  {Emsellem} E.,  {Xu} D.,  {Springel} V.,
  {Krajnovi{\'c}} D.,  2018a, \mn@doi [\mnras] {10.1093/mnras/stx2374}, \href
  {http://adsabs.harvard.edu/abs/2018MNRAS.473.1489L} {473, 1489}

\bibitem[\protect\citeauthoryear{{Li H.} et~al.,}{{Li H.}
  et~al.}{2018b}]{2018MNRAS.476.1765L}
{Li H.} Y.,  et~al., 2018b, \mn@doi [\mnras] {10.1093/mnras/sty334}, \href
  {http://adsabs.harvard.edu/abs/2018MNRAS.476.1765L} {476, 1765}

\bibitem[\protect\citeauthoryear{{Li H.}, {Mao}, {Cappellari}, {Graham},
  {Emsellem}  \& {Long}}{{Li H.} et~al.}{2018c}]{2018ApJ...863L..19L}
{Li H.} Y.,  {Mao} S.,  {Cappellari} M.,  {Graham} M.~T.,  {Emsellem} E.,
  {Long} R.~J.,  2018c, \mn@doi [\apjl] {10.3847/2041-8213/aad54b}, \href
  {http://adsabs.harvard.edu/abs/2018ApJ...863L..19L} {863, L19}

\bibitem[\protect\citeauthoryear{{Li}, {Shu}  \& {Wang}}{{Li}
  et~al.}{2018}]{2018MNRAS.480..431L}
{Li} R.~.,  {Shu} Y.,   {Wang} J.,  2018, \mn@doi [\mnras]
  {10.1093/mnras/sty1813}, \href
  {http://adsabs.harvard.edu/abs/2018MNRAS.480..431L} {480, 431}

\bibitem[\protect\citeauthoryear{{Lin} et~al.,}{{Lin}
  et~al.}{2008}]{2008ApJ...681..232L}
{Lin} L.,  et~al., 2008, \mn@doi [\apj] {10.1086/587928}, \href
  {http://adsabs.harvard.edu/abs/2008ApJ...681..232L} {681, 232}

\bibitem[\protect\citeauthoryear{{Lovell} et~al.,}{{Lovell}
  et~al.}{2018}]{2018MNRAS.481.1950L}
{Lovell} M.~R.,  et~al., 2018, \mn@doi [\mnras] {10.1093/mnras/sty2339}, \href
  {http://adsabs.harvard.edu/abs/2018MNRAS.481.1950L} {481, 1950}

\bibitem[\protect\citeauthoryear{{Lyskova}, {Churazov}  \& {Naab}}{{Lyskova}
  et~al.}{2018}]{2018MNRAS.475.2403L}
{Lyskova} N.,  {Churazov} E.,   {Naab} T.,  2018, \mn@doi [\mnras]
  {10.1093/mnras/sty018}, \href
  {http://adsabs.harvard.edu/abs/2018MNRAS.475.2403L} {475, 2403}

\bibitem[\protect\citeauthoryear{{Mandelbaum}, {Seljak}, {Kauffmann}, {Hirata}
  \& {Brinkmann}}{{Mandelbaum} et~al.}{2006}]{2006MNRAS.368..715M}
{Mandelbaum} R.,  {Seljak} U.,  {Kauffmann} G.,  {Hirata} C.~M.,   {Brinkmann}
  J.,  2006, \mn@doi [\mnras] {10.1111/j.1365-2966.2006.10156.x}, \href
  {http://adsabs.harvard.edu/abs/2006MNRAS.368..715M} {368, 715}

\bibitem[\protect\citeauthoryear{{Marinacci} et~al.,}{{Marinacci}
  et~al.}{2018}]{2018MNRAS.480.5113M}
{Marinacci} F.,  et~al., 2018, \mn@doi [\mnras] {10.1093/mnras/sty2206}, \href
  {http://adsabs.harvard.edu/abs/2018MNRAS.480.5113M} {480, 5113}

\bibitem[\protect\citeauthoryear{{Martizzi}, {Teyssier}  \& {Moore}}{{Martizzi}
  et~al.}{2012a}]{2012MNRAS.420.2859M}
{Martizzi} D.,  {Teyssier} R.,   {Moore} B.,  2012a, \mn@doi [\mnras]
  {10.1111/j.1365-2966.2011.19950.x}, \href
  {https://ui.adsabs.harvard.edu/abs/2012MNRAS.420.2859M} {420, 2859}

\bibitem[\protect\citeauthoryear{{Martizzi}, {Teyssier}, {Moore}  \&
  {Wentz}}{{Martizzi} et~al.}{2012b}]{2012MNRAS.422.3081M}
{Martizzi} D.,  {Teyssier} R.,  {Moore} B.,   {Wentz} T.,  2012b, \mn@doi
  [\mnras] {10.1111/j.1365-2966.2012.20879.x}, \href
  {https://ui.adsabs.harvard.edu/abs/2012MNRAS.422.3081M} {422, 3081}

\bibitem[\protect\citeauthoryear{{Martizzi}, {Teyssier}  \& {Moore}}{{Martizzi}
  et~al.}{2013}]{2013MNRAS.432.1947M}
{Martizzi} D.,  {Teyssier} R.,   {Moore} B.,  2013, \mn@doi [\mnras]
  {10.1093/mnras/stt297}, \href
  {http://adsabs.harvard.edu/abs/2013MNRAS.432.1947M} {432, 1947}

\bibitem[\protect\citeauthoryear{{McConnell} \& {Ma}}{{McConnell} \&
  {Ma}}{2013}]{2013ApJ...764..184M}
{McConnell} N.~J.,  {Ma} C.-P.,  2013, \mn@doi [\apj]
  {10.1088/0004-637X/764/2/184}, \href
  {http://adsabs.harvard.edu/abs/2013ApJ...764..184M} {764, 184}

\bibitem[\protect\citeauthoryear{{Moster}, {Naab}  \& {White}}{{Moster}
  et~al.}{2013}]{2013MNRAS.428.3121M}
{Moster} B.~P.,  {Naab} T.,   {White} S.~D.~M.,  2013, \mn@doi [\mnras]
  {10.1093/mnras/sts261}, \href
  {http://adsabs.harvard.edu/abs/2013MNRAS.428.3121M} {428, 3121}

\bibitem[\protect\citeauthoryear{{Mukherjee}, {Koopmans}, {Metcalf}, {Tortora},
  {Schaller}, {Schaye}, {Vernardos}  \& {Bellagamba}}{{Mukherjee}
  et~al.}{2019}]{2019arXiv190101095M}
{Mukherjee} S.,  {Koopmans} L.~V.~E.,  {Metcalf} R.~B.,  {Tortora} C.,
  {Schaller} M.,  {Schaye} J.,  {Vernardos} G.,   {Bellagamba} F.,  2019, arXiv
  e-prints, \href {http://adsabs.harvard.edu/abs/2019arXiv190101095M} {}

\bibitem[\protect\citeauthoryear{{Naab}, {Johansson}, {Ostriker}  \&
  {Efstathiou}}{{Naab} et~al.}{2007}]{2007ApJ...658..710N}
{Naab} T.,  {Johansson} P.~H.,  {Ostriker} J.~P.,   {Efstathiou} G.,  2007,
  \mn@doi [\apj] {10.1086/510841}, \href
  {http://adsabs.harvard.edu/abs/2007ApJ...658..710N} {658, 710}

\bibitem[\protect\citeauthoryear{{Naab} et~al.,}{{Naab}
  et~al.}{2014}]{2014MNRAS.444.3357N}
{Naab} T.,  et~al., 2014, \mn@doi [\mnras] {10.1093/mnras/stt1919}, \href
  {http://adsabs.harvard.edu/abs/2014MNRAS.444.3357N} {444, 3357}

\bibitem[\protect\citeauthoryear{{Naiman} et~al.,}{{Naiman}
  et~al.}{2018}]{2018MNRAS.477.1206N}
{Naiman} J.~P.,  et~al., 2018, \mn@doi [\mnras] {10.1093/mnras/sty618}, \href
  {http://adsabs.harvard.edu/abs/2018MNRAS.477.1206N} {477, 1206}

\bibitem[\protect\citeauthoryear{{Napolitano}, {Pota}, {Romanowsky}, {Forbes},
  {Brodie}  \& {Foster}}{{Napolitano} et~al.}{2014}]{2014MNRAS.439..659N}
{Napolitano} N.~R.,  {Pota} V.,  {Romanowsky} A.~J.,  {Forbes} D.~A.,  {Brodie}
  J.~P.,   {Foster} C.,  2014, \mn@doi [\mnras] {10.1093/mnras/stt2484}, \href
  {http://adsabs.harvard.edu/abs/2014MNRAS.439..659N} {439, 659}

\bibitem[\protect\citeauthoryear{{Nelson} et~al.,}{{Nelson}
  et~al.}{2015a}]{2015A&C....13...12N}
{Nelson} D.,  et~al., 2015a, \mn@doi [Astronomy and Computing]
  {10.1016/j.ascom.2015.09.003}, \href
  {http://adsabs.harvard.edu/abs/2015A%26C....13...12N} {13, 12}

\bibitem[\protect\citeauthoryear{{Nelson}, {Genel}, {Vogelsberger}, {Springel},
  {Sijacki}, {Torrey}  \& {Hernquist}}{{Nelson}
  et~al.}{2015b}]{2015MNRAS.448...59N}
{Nelson} D.,  {Genel} S.,  {Vogelsberger} M.,  {Springel} V.,  {Sijacki} D.,
  {Torrey} P.,   {Hernquist} L.,  2015b, \mn@doi [\mnras]
  {10.1093/mnras/stv017}, \href
  {http://adsabs.harvard.edu/abs/2015MNRAS.448...59N} {448, 59}

\bibitem[\protect\citeauthoryear{{Nelson} et~al.,}{{Nelson}
  et~al.}{2018}]{2018MNRAS.475..624N}
{Nelson} D.,  et~al., 2018, \mn@doi [\mnras] {10.1093/mnras/stx3040}, \href
  {http://adsabs.harvard.edu/abs/2018MNRAS.475..624N} {475, 624}

\bibitem[\protect\citeauthoryear{{Nelson} et~al.,}{{Nelson}
  et~al.}{2019}]{2019ComAC...6....2N}
{Nelson} D.,  et~al., 2019, \mn@doi [Computational Astrophysics and Cosmology]
  {10.1186/s40668-019-0028-x}, \href
  {https://ui.adsabs.harvard.edu/abs/2019ComAC...6....2N} {6, 2}

\bibitem[\protect\citeauthoryear{{Newman}, {Belli}, {Ellis}  \&
  {Patel}}{{Newman} et~al.}{2018}]{2018ApJ...862..125N}
{Newman} A.~B.,  {Belli} S.,  {Ellis} R.~S.,   {Patel} S.~G.,  2018, \mn@doi
  [\apj] {10.3847/1538-4357/aacd4d}, \href
  {https://ui.adsabs.harvard.edu/abs/2018ApJ...862..125N} {862, 125}

\bibitem[\protect\citeauthoryear{{Nipoti}, {Treu}  \& {Bolton}}{{Nipoti}
  et~al.}{2009a}]{2009ApJ...703.1531N}
{Nipoti} C.,  {Treu} T.,   {Bolton} A.~S.,  2009a, \mn@doi [\apj]
  {10.1088/0004-637X/703/2/1531}, \href
  {http://adsabs.harvard.edu/abs/2009ApJ...703.1531N} {703, 1531}

\bibitem[\protect\citeauthoryear{{Nipoti}, {Treu}, {Auger}  \&
  {Bolton}}{{Nipoti} et~al.}{2009b}]{2009ApJ...706L..86N}
{Nipoti} C.,  {Treu} T.,  {Auger} M.~W.,   {Bolton} A.~S.,  2009b, \mn@doi
  [\apjl] {10.1088/0004-637X/706/1/L86}, \href
  {http://adsabs.harvard.edu/abs/2009ApJ...706L..86N} {706, L86}

\bibitem[\protect\citeauthoryear{{Oser}, {Ostriker}, {Naab}, {Johansson}  \&
  {Burkert}}{{Oser} et~al.}{2010}]{2010ApJ...725.2312O}
{Oser} L.,  {Ostriker} J.~P.,  {Naab} T.,  {Johansson} P.~H.,   {Burkert} A.,
  2010, \mn@doi [\apj] {10.1088/0004-637X/725/2/2312}, \href
  {http://adsabs.harvard.edu/abs/2010ApJ...725.2312O} {725, 2312}

\bibitem[\protect\citeauthoryear{{Peirani}, {Kay}  \& {Silk}}{{Peirani}
  et~al.}{2008}]{2008A&A...479..123P}
{Peirani} S.,  {Kay} S.,   {Silk} J.,  2008, \mn@doi [Astronomy and
  Astrophysics] {10.1051/0004-6361:20077956}, \href
  {https://ui.adsabs.harvard.edu/abs/2008A&A...479..123P} {479, 123}

\bibitem[\protect\citeauthoryear{{Peirani} et~al.,}{{Peirani}
  et~al.}{2017}]{2017MNRAS.472.2153P}
{Peirani} S.,  et~al., 2017, \mn@doi [\mnras] {10.1093/mnras/stx2099}, \href
  {http://adsabs.harvard.edu/abs/2017MNRAS.472.2153P} {472, 2153}

\bibitem[\protect\citeauthoryear{{Peirani} et~al.,}{{Peirani}
  et~al.}{2019}]{2019MNRAS.483.4615P}
{Peirani} S.,  et~al., 2019, \mn@doi [\mnras] {10.1093/mnras/sty3475}, \href
  {http://adsabs.harvard.edu/abs/2019MNRAS.483.4615P} {483, 4615}

\bibitem[\protect\citeauthoryear{{Pillepich} et~al.,}{{Pillepich}
  et~al.}{2018a}]{2018MNRAS.473.4077P}
{Pillepich} A.,  et~al., 2018a, \mn@doi [\mnras] {10.1093/mnras/stx2656}, \href
  {http://adsabs.harvard.edu/abs/2018MNRAS.473.4077P} {473, 4077}

\bibitem[\protect\citeauthoryear{{Pillepich} et~al.,}{{Pillepich}
  et~al.}{2018b}]{2018MNRAS.475..648P}
{Pillepich} A.,  et~al., 2018b, \mn@doi [\mnras] {10.1093/mnras/stx3112}, \href
  {http://adsabs.harvard.edu/abs/2018MNRAS.475..648P} {475, 648}

\bibitem[\protect\citeauthoryear{{Pinkney} et~al.,}{{Pinkney}
  et~al.}{2003}]{2003ApJ...596..903P}
{Pinkney} J.,  et~al., 2003, \mn@doi [\apj] {10.1086/378118}, \href
  {http://adsabs.harvard.edu/abs/2003ApJ...596..903P} {596, 903}

\bibitem[\protect\citeauthoryear{{Planck Collaboration} et~al.,}{{Planck
  Collaboration} et~al.}{2016}]{2016A&A...594A..13P}
{Planck Collaboration} et~al., 2016, \mn@doi [\aap]
  {10.1051/0004-6361/201525830}, \href
  {http://adsabs.harvard.edu/abs/2016A%26A...594A..13P} {594, A13}

\bibitem[\protect\citeauthoryear{{Poci}, {Cappellari}  \& {McDermid}}{{Poci}
  et~al.}{2017}]{2017MNRAS.467.1397P}
{Poci} A.,  {Cappellari} M.,   {McDermid} R.~M.,  2017, \mn@doi [\mnras]
  {10.1093/mnras/stx101}, \href
  {http://adsabs.harvard.edu/abs/2017MNRAS.467.1397P} {467, 1397}

\bibitem[\protect\citeauthoryear{{Rees} \& {Ostriker}}{{Rees} \&
  {Ostriker}}{1977}]{1977MNRAS.179..541R}
{Rees} M.~J.,  {Ostriker} J.~P.,  1977, \mn@doi [\mnras]
  {10.1093/mnras/179.4.541}, \href
  {http://adsabs.harvard.edu/abs/1977MNRAS.179..541R} {179, 541}

\bibitem[\protect\citeauthoryear{{Remus}, {Burkert}, {Dolag}, {Johansson},
  {Naab}, {Oser}  \& {Thomas}}{{Remus} et~al.}{2013}]{2013ApJ...766...71R}
{Remus} R.-S.,  {Burkert} A.,  {Dolag} K.,  {Johansson} P.~H.,  {Naab} T.,
  {Oser} L.,   {Thomas} J.,  2013, \mn@doi [\apj] {10.1088/0004-637X/766/2/71},
  \href {http://adsabs.harvard.edu/abs/2013ApJ...766...71R} {766, 71}

\bibitem[\protect\citeauthoryear{{Remus}, {Dolag}, {Naab}, {Burkert},
  {Hirschmann}, {Hoffmann}  \& {Johansson}}{{Remus}
  et~al.}{2017}]{2017MNRAS.464.3742R}
{Remus} R.-S.,  {Dolag} K.,  {Naab} T.,  {Burkert} A.,  {Hirschmann} M.,
  {Hoffmann} T.~L.,   {Johansson} P.~H.,  2017, \mn@doi [\mnras]
  {10.1093/mnras/stw2594}, \href
  {http://adsabs.harvard.edu/abs/2017MNRAS.464.3742R} {464, 3742}

\bibitem[\protect\citeauthoryear{{Robertson}, {Hernquist}, {Cox}, {Di Matteo},
  {Hopkins}, {Martini}  \& {Springel}}{{Robertson}
  et~al.}{2006}]{2006ApJ...641...90R}
{Robertson} B.,  {Hernquist} L.,  {Cox} T.~J.,  {Di Matteo} T.,  {Hopkins}
  P.~F.,  {Martini} P.,   {Springel} V.,  2006, \mn@doi [\apj]
  {10.1086/500348}, \href {http://adsabs.harvard.edu/abs/2006ApJ...641...90R}
  {641, 90}

\bibitem[\protect\citeauthoryear{{Rodriguez-Gomez} et~al.,}{{Rodriguez-Gomez}
  et~al.}{2015}]{2015MNRAS.449...49R}
{Rodriguez-Gomez} V.,  et~al., 2015, \mn@doi [\mnras] {10.1093/mnras/stv264},
  \href {http://adsabs.harvard.edu/abs/2015MNRAS.449...49R} {449, 49}

\bibitem[\protect\citeauthoryear{{Rodriguez-Gomez} et~al.,}{{Rodriguez-Gomez}
  et~al.}{2016}]{2016MNRAS.458.2371R}
{Rodriguez-Gomez} V.,  et~al., 2016, \mn@doi [\mnras] {10.1093/mnras/stw456},
  \href {http://adsabs.harvard.edu/abs/2016MNRAS.458.2371R} {458, 2371}

\bibitem[\protect\citeauthoryear{{Ruff}, {Gavazzi}, {Marshall}, {Treu}, {Auger}
   \& {Brault}}{{Ruff} et~al.}{2011}]{2011ApJ...727...96R}
{Ruff} A.~J.,  {Gavazzi} R.,  {Marshall} P.~J.,  {Treu} T.,  {Auger} M.~W.,
  {Brault} F.,  2011, \mn@doi [\apj] {10.1088/0004-637X/727/2/96}, \href
  {http://adsabs.harvard.edu/abs/2011ApJ...727...96R} {727, 96}

\bibitem[\protect\citeauthoryear{{Schaye} et~al.,}{{Schaye}
  et~al.}{2015}]{2015MNRAS.446..521S}
{Schaye} J.,  et~al., 2015, \mn@doi [\mnras] {10.1093/mnras/stu2058}, \href
  {http://adsabs.harvard.edu/abs/2015MNRAS.446..521S} {446, 521}

\bibitem[\protect\citeauthoryear{{Serra}, {Oosterloo}, {Cappellari}, {den
  Heijer}  \& {J{\'o}zsa}}{{Serra} et~al.}{2016}]{2016MNRAS.460.1382S}
{Serra} P.,  {Oosterloo} T.,  {Cappellari} M.,  {den Heijer} M.,   {J{\'o}zsa}
  G.~I.~G.,  2016, \mn@doi [\mnras] {10.1093/mnras/stw1010}, \href
  {http://adsabs.harvard.edu/abs/2016MNRAS.460.1382S} {460, 1382}

\bibitem[\protect\citeauthoryear{{Shankar} et~al.,}{{Shankar}
  et~al.}{2017}]{2017ApJ...840...34S}
{Shankar} F.,  et~al., 2017, \mn@doi [\apj] {10.3847/1538-4357/aa66ce}, \href
  {http://adsabs.harvard.edu/abs/2017ApJ...840...34S} {840, 34}

\bibitem[\protect\citeauthoryear{{Shankar} et~al.,}{{Shankar}
  et~al.}{2018}]{2018MNRAS.475.2878S}
{Shankar} F.,  et~al., 2018, \mn@doi [\mnras] {10.1093/mnras/stx3086}, \href
  {http://adsabs.harvard.edu/abs/2018MNRAS.475.2878S} {475, 2878}

\bibitem[\protect\citeauthoryear{{Shu} et~al.,}{{Shu}
  et~al.}{2015}]{2015ApJ...803...71S}
{Shu} Y.,  et~al., 2015, \mn@doi [\apj] {10.1088/0004-637X/803/2/71}, \href
  {http://adsabs.harvard.edu/abs/2015ApJ...803...71S} {803, 71}

\bibitem[\protect\citeauthoryear{{Shu} et~al.,}{{Shu}
  et~al.}{2017}]{2017ApJ...851...48S}
{Shu} Y.,  et~al., 2017, \mn@doi [\apj] {10.3847/1538-4357/aa9794}, \href
  {http://adsabs.harvard.edu/abs/2017ApJ...851...48S} {851, 48}

\bibitem[\protect\citeauthoryear{{Sijacki}, {Springel}, {Di Matteo}  \&
  {Hernquist}}{{Sijacki} et~al.}{2007}]{2007MNRAS.380..877S}
{Sijacki} D.,  {Springel} V.,  {Di Matteo} T.,   {Hernquist} L.,  2007, \mn@doi
  [\mnras] {10.1111/j.1365-2966.2007.12153.x}, \href
  {https://ui.adsabs.harvard.edu/abs/2007MNRAS.380..877S} {380, 877}

\bibitem[\protect\citeauthoryear{{Sijacki}, {Vogelsberger}, {Genel},
  {Springel}, {Torrey}, {Snyder}, {Nelson}  \& {Hernquist}}{{Sijacki}
  et~al.}{2015}]{2015MNRAS.452..575S}
{Sijacki} D.,  {Vogelsberger} M.,  {Genel} S.,  {Springel} V.,  {Torrey} P.,
  {Snyder} G.~F.,  {Nelson} D.,   {Hernquist} L.,  2015, \mn@doi [\mnras]
  {10.1093/mnras/stv1340}, \href
  {http://adsabs.harvard.edu/abs/2015MNRAS.452..575S} {452, 575}

\bibitem[\protect\citeauthoryear{{Silk} \& {Rees}}{{Silk} \&
  {Rees}}{1998}]{1998A&A...331L...1S}
{Silk} J.,  {Rees} M.~J.,  1998, \aap, \href
  {http://adsabs.harvard.edu/abs/1998A%26A...331L...1S} {331, L1}

\bibitem[\protect\citeauthoryear{{Sonnenfeld}, {Treu}, {Gavazzi}, {Suyu},
  {Marshall}, {Auger}  \& {Nipoti}}{{Sonnenfeld}
  et~al.}{2013}]{2013ApJ...777...98S}
{Sonnenfeld} A.,  {Treu} T.,  {Gavazzi} R.,  {Suyu} S.~H.,  {Marshall} P.~J.,
  {Auger} M.~W.,   {Nipoti} C.,  2013, \mn@doi [\apj]
  {10.1088/0004-637X/777/2/98}, \href
  {http://adsabs.harvard.edu/abs/2013ApJ...777...98S} {777, 98}

\bibitem[\protect\citeauthoryear{{Sonnenfeld}, {Nipoti}  \&
  {Treu}}{{Sonnenfeld} et~al.}{2014}]{2014ApJ...786...89S}
{Sonnenfeld} A.,  {Nipoti} C.,   {Treu} T.,  2014, \mn@doi [\apj]
  {10.1088/0004-637X/786/2/89}, \href
  {http://adsabs.harvard.edu/abs/2014ApJ...786...89S} {786, 89}

\bibitem[\protect\citeauthoryear{{Springel}}{{Springel}}{2010}]{2010MNRAS.401..791S}
{Springel} V.,  2010, \mn@doi [\mnras] {10.1111/j.1365-2966.2009.15715.x},
  \href {http://adsabs.harvard.edu/abs/2010MNRAS.401..791S} {401, 791}

\bibitem[\protect\citeauthoryear{{Springel}, {White}, {Tormen}  \&
  {Kauffmann}}{{Springel} et~al.}{2001}]{2001MNRAS.328..726S}
{Springel} V.,  {White} S.~D.~M.,  {Tormen} G.,   {Kauffmann} G.,  2001,
  \mn@doi [\mnras] {10.1046/j.1365-8711.2001.04912.x}, \href
  {http://adsabs.harvard.edu/abs/2001MNRAS.328..726S} {328, 726}

\bibitem[\protect\citeauthoryear{{Springel}, {Di Matteo}  \&
  {Hernquist}}{{Springel} et~al.}{2005}]{2005MNRAS.361..776S}
{Springel} V.,  {Di Matteo} T.,   {Hernquist} L.,  2005, \mn@doi [\mnras]
  {10.1111/j.1365-2966.2005.09238.x}, \href
  {http://adsabs.harvard.edu/abs/2005MNRAS.361..776S} {361, 776}

\bibitem[\protect\citeauthoryear{{Springel} et~al.,}{{Springel}
  et~al.}{2018}]{2018MNRAS.475..676S}
{Springel} V.,  et~al., 2018, \mn@doi [\mnras] {10.1093/mnras/stx3304}, \href
  {http://adsabs.harvard.edu/abs/2018MNRAS.475..676S} {475, 676}

\bibitem[\protect\citeauthoryear{{Tagore}, {Barnes}, {Jackson}, {Kay},
  {Schaller}, {Schaye}  \& {Theuns}}{{Tagore}
  et~al.}{2018}]{2018MNRAS.474.3403T}
{Tagore} A.~S.,  {Barnes} D.~J.,  {Jackson} N.,  {Kay} S.~T.,  {Schaller} M.,
  {Schaye} J.,   {Theuns} T.,  2018, \mn@doi [\mnras] {10.1093/mnras/stx2965},
  \href {http://adsabs.harvard.edu/abs/2018MNRAS.474.3403T} {474, 3403}

\bibitem[\protect\citeauthoryear{{Torrey}, {Vogelsberger}, {Genel}, {Sijacki},
  {Springel}  \& {Hernquist}}{{Torrey} et~al.}{2014}]{2014MNRAS.438.1985T}
{Torrey} P.,  {Vogelsberger} M.,  {Genel} S.,  {Sijacki} D.,  {Springel} V.,
  {Hernquist} L.,  2014, \mn@doi [\mnras] {10.1093/mnras/stt2295}, \href
  {http://adsabs.harvard.edu/abs/2014MNRAS.438.1985T} {438, 1985}

\bibitem[\protect\citeauthoryear{{Torrey} et~al.,}{{Torrey}
  et~al.}{2017}]{2017arXiv171105261T}
{Torrey} P.,  et~al., 2017, preprint, \href
  {http://adsabs.harvard.edu/abs/2017arXiv171105261T} {} (\mn@eprint {arXiv}
  {1711.05261})

\bibitem[\protect\citeauthoryear{{Torrey} et~al.,}{{Torrey}
  et~al.}{2018}]{2018MNRAS.477L..16T}
{Torrey} P.,  et~al., 2018, \mn@doi [\mnras] {10.1093/mnrasl/sly031}, \href
  {http://adsabs.harvard.edu/abs/2018MNRAS.477L..16T} {477, L16}

\bibitem[\protect\citeauthoryear{{Tortora}, {La Barbera}, {Napolitano},
  {Romanowsky}, {Ferreras}  \& {de Carvalho}}{{Tortora}
  et~al.}{2014a}]{2014MNRAS.445..115T}
{Tortora} C.,  {La Barbera} F.,  {Napolitano} N.~R.,  {Romanowsky} A.~J.,
  {Ferreras} I.,   {de Carvalho} R.~R.,  2014a, \mn@doi [\mnras]
  {10.1093/mnras/stu1616}, \href
  {http://adsabs.harvard.edu/abs/2014MNRAS.445..115T} {445, 115}

\bibitem[\protect\citeauthoryear{{Tortora}, {Napolitano}, {Saglia},
  {Romanowsky}, {Covone}  \& {Capaccioli}}{{Tortora}
  et~al.}{2014b}]{2014MNRAS.445..162T}
{Tortora} C.,  {Napolitano} N.~R.,  {Saglia} R.~P.,  {Romanowsky} A.~J.,
  {Covone} G.,   {Capaccioli} M.,  2014b, \mn@doi [\mnras]
  {10.1093/mnras/stu1712}, \href
  {http://adsabs.harvard.edu/abs/2014MNRAS.445..162T} {445, 162}

\bibitem[\protect\citeauthoryear{{Veale} et~al.,}{{Veale}
  et~al.}{2017}]{2017MNRAS.464..356V}
{Veale} M.,  et~al., 2017, \mn@doi [\mnras] {10.1093/mnras/stw2330}, \href
  {http://adsabs.harvard.edu/abs/2017MNRAS.464..356V} {464, 356}

\bibitem[\protect\citeauthoryear{{Vogelsberger}, {Genel}, {Sijacki}, {Torrey},
  {Springel}  \& {Hernquist}}{{Vogelsberger}
  et~al.}{2013}]{2013MNRAS.436.3031V}
{Vogelsberger} M.,  {Genel} S.,  {Sijacki} D.,  {Torrey} P.,  {Springel} V.,
  {Hernquist} L.,  2013, \mn@doi [\mnras] {10.1093/mnras/stt1789}, \href
  {http://adsabs.harvard.edu/abs/2013MNRAS.436.3031V} {436, 3031}

\bibitem[\protect\citeauthoryear{{Vogelsberger} et~al.,}{{Vogelsberger}
  et~al.}{2014a}]{2014MNRAS.444.1518V}
{Vogelsberger} M.,  et~al., 2014a, \mn@doi [\mnras] {10.1093/mnras/stu1536},
  \href {http://adsabs.harvard.edu/abs/2014MNRAS.444.1518V} {444, 1518}

\bibitem[\protect\citeauthoryear{{Vogelsberger} et~al.,}{{Vogelsberger}
  et~al.}{2014b}]{2014Natur.509..177V}
{Vogelsberger} M.,  et~al., 2014b, \mn@doi [\nat] {10.1038/nature13316}, \href
  {http://adsabs.harvard.edu/abs/2014Natur.509..177V} {509, 177}

\bibitem[\protect\citeauthoryear{{Vogelsberger} et~al.,}{{Vogelsberger}
  et~al.}{2018}]{2018MNRAS.474.2073V}
{Vogelsberger} M.,  et~al., 2018, \mn@doi [\mnras] {10.1093/mnras/stx2955},
  \href {http://adsabs.harvard.edu/abs/2018MNRAS.474.2073V} {474, 2073}

\bibitem[\protect\citeauthoryear{{Vogelsberger}, {Marinacci}, {Torrey}  \&
  {Puchwein}}{{Vogelsberger} et~al.}{2019a}]{2019arXiv190907976V}
{Vogelsberger} M.,  {Marinacci} F.,  {Torrey} P.,   {Puchwein} E.,  2019a,
  arXiv e-prints, \href {https://ui.adsabs.harvard.edu/abs/2019arXiv190907976V}
  {}

\bibitem[\protect\citeauthoryear{{Vogelsberger} et~al.,}{{Vogelsberger}
  et~al.}{2019b}]{2019arXiv190407238V}
{Vogelsberger} M.,  et~al., 2019b, arXiv e-prints, \href
  {http://adsabs.harvard.edu/abs/2019arXiv190407238V} {}

\bibitem[\protect\citeauthoryear{{Wang} et~al.,}{{Wang}
  et~al.}{2018}]{2018arXiv181106545W}
{Wang} Y.,  et~al., 2018, preprint, \href
  {http://adsabs.harvard.edu/abs/2018arXiv181106545W} {} (\mn@eprint {arXiv}
  {1811.06545})

\bibitem[\protect\citeauthoryear{{Weinberger} et~al.,}{{Weinberger}
  et~al.}{2017}]{2017MNRAS.465.3291W}
{Weinberger} R.,  et~al., 2017, \mn@doi [\mnras] {10.1093/mnras/stw2944}, \href
  {http://adsabs.harvard.edu/abs/2017MNRAS.465.3291W} {465, 3291}

\bibitem[\protect\citeauthoryear{{Weinberger} et~al.,}{{Weinberger}
  et~al.}{2018}]{2018MNRAS.479.4056W}
{Weinberger} R.,  et~al., 2018, \mn@doi [\mnras] {10.1093/mnras/sty1733}, \href
  {http://adsabs.harvard.edu/abs/2018MNRAS.479.4056W} {479, 4056}

\bibitem[\protect\citeauthoryear{{Wellons} et~al.,}{{Wellons}
  et~al.}{2015}]{2015MNRAS.449..361W}
{Wellons} S.,  et~al., 2015, \mn@doi [\mnras] {10.1093/mnras/stv303}, \href
  {http://adsabs.harvard.edu/abs/2015MNRAS.449..361W} {449, 361}

\bibitem[\protect\citeauthoryear{{Wellons} et~al.,}{{Wellons}
  et~al.}{2016}]{2016MNRAS.456.1030W}
{Wellons} S.,  et~al., 2016, \mn@doi [\mnras] {10.1093/mnras/stv2738}, \href
  {http://adsabs.harvard.edu/abs/2016MNRAS.456.1030W} {456, 1030}

\bibitem[\protect\citeauthoryear{{Whitaker} et~al.,}{{Whitaker}
  et~al.}{2017}]{2017ApJ...838...19W}
{Whitaker} K.~E.,  et~al., 2017, \mn@doi [\apj] {10.3847/1538-4357/aa6258},
  \href {http://adsabs.harvard.edu/abs/2017ApJ...838...19W} {838, 19}

\bibitem[\protect\citeauthoryear{{White} \& {Frenk}}{{White} \&
  {Frenk}}{1991}]{1991ApJ...379...52W}
{White} S.~D.~M.,  {Frenk} C.~S.,  1991, \mn@doi [\apj] {10.1086/170483}, \href
  {http://adsabs.harvard.edu/abs/1991ApJ...379...52W} {379, 52}

\bibitem[\protect\citeauthoryear{{White} \& {Rees}}{{White} \&
  {Rees}}{1978}]{1978MNRAS.183..341W}
{White} S.~D.~M.,  {Rees} M.~J.,  1978, \mn@doi [\mnras]
  {10.1093/mnras/183.3.341}, \href
  {http://adsabs.harvard.edu/abs/1978MNRAS.183..341W} {183, 341}

\bibitem[\protect\citeauthoryear{{Woo}, {Dekel}, {Faber}  \& {Koo}}{{Woo}
  et~al.}{2015}]{2015MNRAS.448..237W}
{Woo} J.,  {Dekel} A.,  {Faber} S.~M.,   {Koo} D.~C.,  2015, \mn@doi [\mnras]
  {10.1093/mnras/stu2755}, \href
  {http://adsabs.harvard.edu/abs/2015MNRAS.448..237W} {448, 237}

\bibitem[\protect\citeauthoryear{{Wyithe} \& {Loeb}}{{Wyithe} \&
  {Loeb}}{2003}]{2003ApJ...595..614W}
{Wyithe} J.~S.~B.,  {Loeb} A.,  2003, \mn@doi [\apj] {10.1086/377475}, \href
  {http://adsabs.harvard.edu/abs/2003ApJ...595..614W} {595, 614}

\bibitem[\protect\citeauthoryear{{Xu}, {Sluse}, {Schneider}, {Springel},
  {Vogelsberger}, {Nelson}  \& {Hernquist}}{{Xu}
  et~al.}{2016}]{2016MNRAS.456..739X}
{Xu} D.,  {Sluse} D.,  {Schneider} P.,  {Springel} V.,  {Vogelsberger} M.,
  {Nelson} D.,   {Hernquist} L.,  2016, \mn@doi [\mnras]
  {10.1093/mnras/stv2708}, \href
  {http://adsabs.harvard.edu/abs/2016MNRAS.456..739X} {456, 739}

\bibitem[\protect\citeauthoryear{{Xu}, {Springel}, {Sluse}, {Schneider},
  {Sonnenfeld}, {Nelson}, {Vogelsberger}  \& {Hernquist}}{{Xu}
  et~al.}{2017}]{2017MNRAS.469.1824X}
{Xu} D.,  {Springel} V.,  {Sluse} D.,  {Schneider} P.,  {Sonnenfeld} A.,
  {Nelson} D.,  {Vogelsberger} M.,   {Hernquist} L.,  2017, \mn@doi [\mnras]
  {10.1093/mnras/stx899}, \href
  {http://adsabs.harvard.edu/abs/2017MNRAS.469.1824X} {469, 1824}

\bibitem[\protect\citeauthoryear{{Yoon}, {Yuan}, {Ostriker}, {Ciotti}  \&
  {Zhu}}{{Yoon} et~al.}{2019}]{2019arXiv190107570Y}
{Yoon} D.,  {Yuan} F.,  {Ostriker} J.~P.,  {Ciotti} L.,   {Zhu} B.,  2019,
  arXiv e-prints, \href {https://ui.adsabs.harvard.edu/abs/2019arXiv190107570Y}
  {p. arXiv:1901.07570}

\bibitem[\protect\citeauthoryear{{Yuan}, {Yoon}, {Li}, {Gan}, {Ho}  \&
  {Guo}}{{Yuan} et~al.}{2018}]{2018ApJ...857..121Y}
{Yuan} F.,  {Yoon} D.,  {Li} Y.-P.,  {Gan} Z.-M.,  {Ho} L.~C.,   {Guo} F.,
  2018, \mn@doi [\apj] {10.3847/1538-4357/aab8f8}, \href
  {https://ui.adsabs.harvard.edu/abs/2018ApJ...857..121Y} {857, 121}

\bibitem[\protect\citeauthoryear{{Zolotov} et~al.,}{{Zolotov}
  et~al.}{2015}]{2015MNRAS.450.2327Z}
{Zolotov} A.,  et~al., 2015, \mn@doi [\mnras] {10.1093/mnras/stv740}, \href
  {http://adsabs.harvard.edu/abs/2015MNRAS.450.2327Z} {450, 2327}

\bibitem[\protect\citeauthoryear{{van Dokkum} et~al.,}{{van Dokkum}
  et~al.}{2015}]{2015ApJ...813...23V}
{van Dokkum} P.~G.,  et~al., 2015, \mn@doi [\apj] {10.1088/0004-637X/813/1/23},
  \href {http://adsabs.harvard.edu/abs/2015ApJ...813...23V} {813, 23}

\bibitem[\protect\citeauthoryear{{van de Voort}, {Schaye}, {Booth}, {Haas}  \&
  {Dalla Vecchia}}{{van de Voort} et~al.}{2011}]{2011MNRAS.414.2458V}
{van de Voort} F.,  {Schaye} J.,  {Booth} C.~M.,  {Haas} M.~R.,   {Dalla
  Vecchia} C.,  2011, \mn@doi [\mnras] {10.1111/j.1365-2966.2011.18565.x},
  \href {http://adsabs.harvard.edu/abs/2011MNRAS.414.2458V} {414, 2458}

\makeatother
\end{thebibliography}


 
\appendix

\section{Statistical analyses for galaxy mergers' impact on $\gamma^{\prime}$}
\label{sec: A}

\begin{table}
		\begin{center}
		\begin{tabular}{llcccc}
			\hline \hline
			Quantity 1 & Quantity 2 & $z$ & $r_{\mathrm{p}}$ & $r_{\mathrm{boot}}$ & $\sigma_{\mathrm{boot}}$ \\
			\hline
			$\mu_{\ast}$ & $\delta\gamma^{\prime}$ & $[0, 1]$ & $-0.143$ & $-0.144$ & $0.036$ \\
			$\mu_{\ast}$ & $\delta\gamma^{\prime}$ & $[1, 2]$ & $-0.195$ & $-0.193$ & $0.044$ \\
			$\mu_{\ast}$ & $\delta\gamma^{\prime}$ & $[2, 4]$ & $-0.133$ & $-0.133$ & $0.042$ \\
			\hline
			$f_{\mathrm{Cold}}$ & $\delta\gamma^{\prime}$ & $[0, 1]$ & $0.212$ & $0.216$ & $0.045$ \\
			$f_{\mathrm{Cold}}$ & $\delta\gamma^{\prime}$ & $[1, 2]$ & $0.317$ & $0.316$ & $0.042$ \\
			$f_{\mathrm{Cold}}$ & $\delta\gamma^{\prime}$ & $[2, 4]$ & $0.203$ & $0.203$ & $0.034$ \\
			\hline
			$\mu_{\ast}$ & $f_{\mathrm{Cold}}$ & $[0, 1]$ & $-0.083$ & $-0.083$ & $0.039$ \\
			$\mu_{\ast}$ & $f_{\mathrm{Cold}}$ & $[1, 2]$ & $-0.127$ & $-0.129$ & $0.046$ \\
			$\mu_{\ast}$ & $f_{\mathrm{Cold}}$ & $[2, 4]$ & $-0.062$ & $-0.063$ & $0.036$ \\
			\hline
		\end{tabular}
        \end{center}
		\caption{Correlations of merger-related quantities chosen from $\mu_{\ast}$, $f_{\mathrm{Cold}}$, and $\delta\gamma^{\prime}$. For each `Quantity 1$-$Quantity 2 correlation', we calculate its Pearson correlation coefficient $r_{\mathrm{p}}$ in the 3 redshift bins $z\in[0,1], [1,2],\ \mathrm{and}\ [2,4]$. For each correlation, we calculate the bootstrap mean $r_{\mathrm{boot}}$ and variance $\sigma_{\mathrm{boot}}$ for its $r_{\mathrm{p}}$ with 1000 bootstrap realizations. The results are consistent with the merger statistics shown in Fig.~\ref{fig:merger}.}
		\label{tab:mu_fCG_corr}
\end{table}

\begin{table*}
		\begin{center}
		\begin{tabular}{llcc}
			\hline \hline
			Quantity 1 & Quantity 2 & $z$ & $p$\\
			\hline
			  &  & $[0, 1]$ & $9\times10^{-4}$ \\
			$\mu_{\ast}\,(\delta\gamma^{\prime} < -0.1)$ & $\mu_{\ast}\,(\delta\gamma^{\prime} > 0.1)$ & $[1, 2]$ & $8\times10^{-6}$ \\
			  &  & [2, 4] & $6\times10^{-5}$ \\
			  \hline
			  &  & $[0, 1]$ & $0.11$ \\
			$\mu_{\ast}\,(\delta\gamma^{\prime} > 0.1)$ & $\mu_{\ast}\,(\delta\gamma^{\prime}\in[-0.1, 0.1])$ & $[1, 2]$ & $0.43$ \\
              &  & $[2, 4]$ & $0.51$ \\
              \hline
              &  & $[0, 1]$ & $2\times10^{-3}$ \\
            $\mu_{\ast}\,(\delta\gamma^{\prime}\in[-0.1, 0.1])$ & $\mu_{\ast}\,(\delta\gamma^{\prime} < -0.1)$ & $[1, 2]$ & $1\times10^{-3}$ \\
              &  & $[2, 4]$ & $2\times10^{-5}$ \\
              
			\hline
			
			  &  & $[0, 1]$ & $7\times10^{-3}$ \\
			$f_{\mathrm{Cold}}\,(\delta\gamma^{\prime} < -0.1)$ & $f_{\mathrm{Cold}}\,(\delta\gamma^{\prime} > 0.1)$ & $[1, 2]$ & $3\times10^{-4}$ \\
			  &  & $[2, 4]$ & $6\times10^{-8}$ \\
			  \hline
			  &  & $[0, 1]$ & $0.03$ \\
			$f_{\mathrm{Cold}}\,(\delta\gamma^{\prime} > 0.1)$ & $f_{\mathrm{Cold}}\,(\delta\gamma^{\prime}\in[-0.1, 0.1])$ & $[1, 2]$ & $3\times10^{-4}$ \\
			  &  & $[2, 4]$ & $3\times10^{-10}$ \\
			  \hline
			  &  & $[0, 1]$ & $2\times 10^{-7}$ \\  
			$f_{\mathrm{Cold}}\,(\delta\gamma^{\prime}\in[-0.1, 0.1])$ & $f_{\mathrm{Cold}}\,(\delta\gamma^{\prime} < -0.1)$ & $[1, 2]$ & $0.44$ \\
			  &  & $[2, 4]$ & $0.42$ \\
			\hline
		\end{tabular}
        \end{center}
		\caption{K-S test $p-$values for merger-related quantities. We carry out the K-S test for the projected distributions of $\mu_{\ast}$ and $f_{\mathrm{Cold}}$ inducing shallower ($\delta\gamma^{\prime}<-0.1$), steeper ($\delta\gamma^{\prime} > 0.1$), and near-constant ($\delta\gamma^{\prime}\in[-0.1, 0.1]$) total power-law density slopes. The K-S test for each pair of merger-related quantities is conducted in three redshift bins, i.e. $[0, 1]$, $[1, 2]$, and $[2, 4]$.}
		\label{tab:K-S_test1}
\end{table*}

To further quantify the impact of mergers on the change in total density profile, we calculate the Pearson correlation coefficient $r_{\mathrm{p}}$ in between merger-related quantities chosen from $\mu_{\ast}$, $f_{\mathrm{Cold}}$, and $\delta\gamma^{\prime}$ in the 3 redshift bins. We also calculate the bootstrap mean $r_{\mathrm{boot}}$ and variance $\sigma_{\mathrm{boot}}$ for $r_{\mathrm{p}}$ in each correlation with 1000 bootstrap realizations. The results of the correlation coefficients and their bootstrap errors are summarized in Table~\ref{tab:mu_fCG_corr}. We show the Kolmogorov-Smirnov (K-S, hereafter) test $p-$values for the projected distributions of $\mu_{\ast}$ and $f_{\mathrm{Cold}}$ inducing different changes of the total power-law density slope in Table~\ref{tab:K-S_test1}. 

A mild negative $\mu_{\ast}-\delta\gamma^{\prime}$ and a mild positive $f_{\mathrm{Cold}}-\delta\gamma^{\prime}$ correlation are shown, which is consistent with the merger statistics in Fig.~\ref{fig:merger}. These correlations are significant, although weak, as shown by the $r_{\mathrm{boot}}$ and $\sigma_{\mathrm{boot}}$ values from bootstrapping which are consistent with $r_{\mathrm{p}} \neq 0$ (Table~\ref{tab:mu_fCG_corr}). However, the $f_{\mathrm{Cold}}-\delta\gamma^{\prime}$ correlation is stronger than the $\mu_{\ast}-\delta\gamma^{\prime}$ correlation, rendering `wet' and `dry' a more decisive factor in affecting the galaxy total density profile compared to the merger mass ratio. This is also reflected in the K-S test $p-$values for $f_{\mathrm{Cold}}$ and $\mu_{\ast}$. $f_{\mathrm{Cold}}(\delta\gamma^{\prime} > 0.1)$ is clearly drawn from a different distribution than $f_{\mathrm{Cold}}(\delta\gamma^{\prime}<-0.1)$ and $f_{\mathrm{Cold}}(\delta\gamma^{\prime}\in[-0.1, 0.1])$ above $z=1$, meaning that the `wet' and `dry' mergers affect the total density profile differently. The dominant `dry' mergers making the total density profile shallower makes $f_{\mathrm{Cold}}(\delta\gamma^{\prime}<-0.1)$ distinct from $f_{\mathrm{Cold}}(\delta\gamma^{\prime}>0.1)$ and $f_{\mathrm{Cold}}(\delta\gamma^{\prime}\in[-0.1, 0.1])$ below $z=1$. Galaxies with larger $\mu_{\ast}$ have more massive infall progenitors that in general are quenched earlier and faster~\citep{2015MNRAS.450.2327Z}. This leads to lower cold gas fraction at the time of the merger resulting in a weak negative $\mu_{\ast}-f_{\mathrm{Cold}}$ correlation (consistent with non-zero $r_{\mathrm{p}}$ from the bootstrap errors). The $p-$values in different $\mu_{\ast}$ distributions also reflect this trend. $\mu_{\ast}(\delta\gamma^{\prime}<-0.1)$ is clearly drawn from a different sample than $\mu_{\ast}(\delta\gamma^{\prime}>0.1)$ and $\mu_{\ast}(\delta\gamma\in[-0.1, 0.1])$ at all redshifts, indicating that `dry' major mergers with higher $\mu_{\ast}$ induce shallower $\gamma^{\prime}$ while gas-rich mergers that cause constant to steeper slopes mainly have lower $\mu_{\ast}$. Hence, the $f_{\mathrm{Cold}}-\delta\gamma^{\prime}$ correlation and the $\mu_{\ast}-f_{\mathrm{Cold}}$ correlation roughly account for the mild $\mu_{\ast}-\delta\gamma^{\prime}$ correlation, suggesting that merger mass ratio is sub-dominant compared with cold gas fraction in altering $\gamma^{\prime}$.

\section{Statistical analyses for AGN feedback's impact on $\gamma^{\prime}$}
\label{sec: B}

\begin{table}
	\begin{center}
	\begin{tabular}{llcccc}
		\hline \hline
		Quantity 1 & Quantity 2 & $z$ & $r_{\mathrm{p}}$ & $r_{\mathrm{boot}}$ & $\sigma_{\mathrm{boot}}$\\
		\hline
		$\Delta E_{\mathrm{Kin}}$ & $\Delta\gamma^{\prime}$ & $[0, 0.5]$ & $-0.023$ & $-0.020$ & $0.066$ \\
		$\Delta E_{\mathrm{Kin}}$ & $\Delta\gamma^{\prime}$ & $[0.5, 1]$ & $0.106$ & $0.101$ & $0.070$ \\
		$\Delta E_{\mathrm{Kin}}$ & $\Delta\gamma^{\prime}$ & $[1, 2.5]$ & $-0.271$ & $-0.281$ & $0.048$ \\
		\hline
		$\Delta E_{\mathrm{Thm}}$ & $\Delta\gamma^{\prime}$ & $[0, 0.5]$ & $-0.058$ & $-0.055$ & $0.090$ \\
		$\Delta E_{\mathrm{Thm}}$ & $\Delta\gamma^{\prime}$ & $[0.5, 1]$ & $0.050$ & $0.032$ & $0.094$ \\
		$\Delta E_{\mathrm{Thm}}$ & $\Delta\gamma^{\prime}$ & $[1, 2.5]$ & $-0.227$ & $-0.234$ & $0.050$ \\
		\hline
	\end{tabular}
    \end{center}
	\caption{Correlations of AGN feedback-related quantities chosen from $\Delta E_{\mathrm{Kin}}$, $\Delta E_{\mathrm{Thm}}$, and $\Delta\gamma^{\prime}$. For each `Quantity 1$-$Quantity 2 correlation', we calculate its Pearson correlation coefficient $r_{\mathrm{p}}$ in the 3 redshift bins $z\in[0,1], [1,2],\mathrm{and}\ [2,4]$. For each correlation, we calculate the bootstrap mean $r_{\mathrm{boot}}$ and variance $\sigma_{\mathrm{boot}}$ for its $r_{\mathrm{p}}$ with 1000 bootstrap realizations. The results are consistent with the AGN feedback energy analysis shown in Fig.~\ref{fig:bh_tracks}.}
	\label{tab:rm_qm_corr}
\end{table}

\begin{table*}
		\begin{center}
		\begin{tabular}{llcc}
			\hline \hline
			Quantity 1 & Quantity 2 & $z$ & $p$\\
			\hline
			  &  & $[0, 0.5]$ & $0.95$ \\
			$\Delta E_{\mathrm{Thm}}\,(\Delta\gamma^{\prime} < -0.1)$ & $\Delta E_{\mathrm{Thm}}\,(\Delta\gamma^{\prime} > 0.1)$ & $[0.5, 1]$ & $0.95$ \\
			  &  & [1, 2.5] & $3\times10^{-5}$ \\
			  \hline
			  &  & $[0, 0.5]$ & $4\times10^{-5}$ \\
			$\Delta E_{\mathrm{Thm}}\,(\Delta\gamma^{\prime} > 0.1)$ & $\Delta E_{\mathrm{Thm}}\,(\Delta\gamma^{\prime}\in[-0.1, 0.1])$ & $[0.5, 1]$ & $0.07$ \\
              &  & $[1, 2.5]$ & $2\times10^{-3}$ \\
              \hline
              &  & $[0, 0.5]$ & $1\times10^{-6}$ \\
            $\Delta E_{\mathrm{Thm}}\,(\Delta\gamma^{\prime}\in[-0.1, 0.1])$ & $\Delta E_{\mathrm{Thm}}\,(\Delta\gamma^{\prime} < -0.1)$ & $[0.5, 1]$ & $2\times10^{-3}$ \\
              &  & $[1, 2.5]$ & $0.52$ \\
              
			\hline
			
			  &  & $[0, 0.5]$ & $0.88$ \\
			$\Delta E_{\mathrm{Kin}}\,(\Delta\gamma^{\prime} < -0.1)$ & $\Delta E_{\mathrm{Kin}}\,(\Delta\gamma^{\prime} > 0.1)$ & $[0.5, 1]$ & $0.34$ \\
			  &  & $[1, 2.5]$ & $2\times10^{-24}$ \\
			  \hline
			  &  & $[0, 0.5]$ & $8\times10^{-4}$ \\
			$\Delta E_{\mathrm{Kin}}\,(\Delta\gamma^{\prime} > 0.1)$ & $\Delta E_{\mathrm{Kin}}\,(\Delta\gamma^{\prime}\in[-0.1, 0.1])$ & $[0.5, 1]$ & $0.04$ \\
			  &  & $[1, 2.5]$ & $2\times10^{-16}$ \\
			  \hline
			  &  & $[0, 0.5]$ & $2\times 10^{-5}$ \\  
			$\Delta E_{\mathrm{Kin}}\,(\Delta\gamma^{\prime}\in[-0.1, 0.1])$ & $\Delta E_{\mathrm{Kin}}\,(\Delta\gamma^{\prime} < -0.1)$ & $[0.5, 1]$ & $8\times10^{-3}$ \\
			  &  & $[1, 2.5]$ & $0.20$ \\
			\hline
		\end{tabular}
        \end{center}
		\caption{K-S test $p-$values for AGN feedback-related quantities. We carry out the K-S test for the projected distributions of $\Delta E_{\mathrm{Thm}}$ and $\Delta E_{\mathrm{Kin}}$ inducing shallower  ($\delta\gamma^{\prime}<-0.1$), steeper ($\delta\gamma^{\prime} > 0.1$), and near-constant ($\delta\gamma^{\prime}\in[-0.1, 0.1]$) total power-law density slopes. The K-S test for each pair of AGN feedback-related quantities is conducted in three redshift bins, i.e. $[0, 0.5]$, $[0.5, 1]$, and $[1, 2.5]$.}
		\label{tab:K-S_test2}
\end{table*}

To further quantify the impact of AGN feedback energy on the change in total density profile, we calculate the Pearson correlation coefficient $r_{\mathrm{p}}$ in between feedback energy-related quantities chosen from $\Delta E_{\mathrm{Kin}}$, $\Delta E_{\mathrm{Thm}}$, and $\Delta\gamma^{\prime}$ in the 3 redshift bins. We also calculate the bootstrap mean $r_{\mathrm{boot}}$ and error $\sigma_{\mathrm{boot}}$ for $r_{\mathrm{p}}$ in each correlation with 1000 bootstrap realizations. The results of the correlation coefficients and their bootstrap errors are summarized in Table~\ref{tab:rm_qm_corr}. We show the K-S test $p-$values for the projected distributions of $\Delta E_{\mathrm{Thm}}$ and $\Delta E_{\mathrm{Kin}}$ inducing different changes of the total power-law density slope in Table~\ref{tab:K-S_test2}. 

Negative $\Delta\gamma^{\prime}-\Delta E_{\mathrm{Kin}}$ and $\Delta\gamma^{\prime}-\Delta E_{\mathrm{Thm}}$ correlations during $z\in[1, 2.5]$ are consistent with the feedback statistics in Fig.~\ref{fig:bh_tracks}. The correlations during $z\in[1, 2.5]$ are significant as shown by the $r_{\mathrm{boot}}$ and $\sigma_{\mathrm{boot}}$ values from bootstrapping which are consistent with $r_{\mathrm{p}}\neq 0$ (Table.~\ref{tab:rm_qm_corr}). However, the absolute value of $r_{\mathrm{p}}$ in the $\Delta\gamma^{\prime}-\Delta E_{\mathrm{Kin}}$ correlation is larger than that in the $\Delta\gamma^{\prime}-\Delta E_{\mathrm{Thm}}$ correlation, indicating that the kinetic mode feedback is more efficient at altering the total density profile, although its feedback energy rate is lower compared to the thermal mode. This is also reflected through the $p-$values in the K-S test. $\Delta E_{\mathrm{Thm}}(\Delta\gamma^{\prime}> 0.1)$ cannot be drawn from the same sample as $\Delta E_{\mathrm{Thm}}(\Delta\gamma^{\prime}<-0.1)$ and $\Delta E_{\mathrm{Thm}}(\Delta\gamma^{\prime}\in[-0.1, 0.1])$, and $\Delta E_{\mathrm{Kin}}(\Delta\gamma^{\prime}> 0.1)$ cannot be drawn from the same sample as $\Delta E_{\mathrm{Kin}}(\Delta\gamma^{\prime}<-0.1)$ and $\Delta E_{\mathrm{Kin}}(\Delta\gamma^{\prime}\in[-0.1, 0.1])$. The $p-$values for the kinetic mode are much smaller than those of the thermal mode, indicating a more distinct impact on the change of the density profile for higher and lower kinetic mode AGN feedback energy. Furthermore, the correlation coefficients for $\Delta\gamma^{\prime}-\Delta E_{\mathrm{Kin}}$ and $\Delta\gamma^{\prime}-\Delta E_{\mathrm{Thm}}$ have a much smaller absolute value during $z\leqslant 1$ (corresponding to the upper two panels, which is consistent with the weaker correlations between feedback energy and change in the total density profile shown in the upper two panels of Fig.~\ref{fig:bh_tracks}. The $r_{\mathrm{p}}$ values are really close to zero for these correaltions, and $r_{\mathrm{boot}}$ and $\sigma_{\mathrm{boot}}$ values from bootstrapping are in fact consistent with no correlation (weak and insignificant, Table~\ref{tab:rm_qm_corr}). From the K-S test $p-$values for both the kinetic and the thermal modes, $\Delta E_{\mathrm{Thm,Kin}}(\Delta\gamma^{\prime}<-0.1)$ and $\Delta E_{\mathrm{Thm,Kin}}(\Delta\gamma^{\prime}>0.1)$ can be drawn from the same distribution below $z=1$, while $\Delta E_{\mathrm{Thm, Kin}}(\Delta\gamma^{\prime}\in[-0.1, 0.1])$ can be drawn from a distinct distribution with the former two quantities, indicating that AGN feedback energy is consistent with inducing constant slopes below $z=1$. 


\bsp	
\label{lastpage}
\end{document}